\documentclass[11pt,a4paper]{article}
\usepackage[utf8]{inputenc}
\usepackage{amsmath}
\usepackage{placeins}
\usepackage{url}
\usepackage{natbib}
\usepackage{color,xcolor,setspace,geometry}
\setcitestyle{authoryear,open={(},close={)}}
\usepackage{cite}
\usepackage{amsfonts}
\usepackage[normalem]{ulem}
\usepackage{pifont}

\usepackage{arydshln}
\usepackage{tikz}
\usetikzlibrary{shapes}
\usetikzlibrary{plotmarks}
\usetikzlibrary{tikzmark,matrix,calc}
\usetikzlibrary{arrows.meta,calc,decorations.markings,math,arrows.meta}
\usepackage{amssymb}
\usepackage{multirow}
\usepackage{graphicx}
\usepackage{soul}
\soulregister\citep7
\soulregister\cite7
\soulregister\citet7
\soulregister\citealp7
\soulregister\ref7
\usepackage{rotating}
\usepackage{setspace}
\usepackage{threeparttable}
\usepackage{authblk}
\usepackage{subfigure}
\usepackage{tikz}

\usepackage{appendix}

\begin{document}
\begin{spacing}{1.5}	

\title{Integrated storage assignment for an e-grocery fulfilment centre: Accounting for day-of-week demand patterns}

\noindent

\author[$\star$ $\dag$]{David Winkelmann}
\author[$\star$]{Frederik Tolkmitt}
\author[$\star$]{Matthias Ulrich}
\author[$\star$]{Michael Römer}

\affil[$\dag$]{Corresponding author: david.winkelmann@uni-bielefeld.de}
\affil[$\star$]{Bielefeld University, Universit\"atsstrasse 25, Bielefeld, Germany.}

\maketitle
\noindent
\vspace{-1.5cm}

\begin{abstract}
In this paper, we deal with a storage assignment problem arising in a fulfilment centre of a major European e-grocery retailer. The centre can be characterised as a hybrid warehouse consisting of a highly efficient and partially automated fast-picking area designed as a pick-and-pass system with multiple stations, and a picker-to-parts area. The storage assignment problem considered in this paper comprises the decisions to select the products to be allocated to the fast-picking area, the assignment of the products to picking stations and the determination of a shelf within the assigned station. The objective is to achieve a high level of picking efficiency while respecting station workload balancing and precedence order constraints. We propose to solve this three-level problem using an integrated MILP model. In computational experiments with real-world data, we show that using the proposed integrated approach yields significantly better results than a sequential approach in which the selection of products to be included in the fast-picking area is solved before assigning station and shelf. Furthermore, we provide an extension to the integrated storage assignment model that explicitly accounts for within-week demand variation. In a set of experiments with day-of-week-dependent demands we show that while a storage assignment that is based on average demand figures tends to exhibit a highly imbalanced workload on certain days of the week, the augmented model yields storage assignments that are well balanced on each day of the week without compromising the quality of the solutions in terms of picking efficiency.

\end{abstract}

\section{Introduction}\label{sec1}

In e-grocery retailing, grocery products are ordered online and directly delivered at a date and time period determined by the customer. In the recent years, the e-grocery business experienced phases of dynamic growth: For example, following \citet{Hubner2019}, online sales of food and beverages were expected to grow by 10\% to 18\% on average per year in Europe, the US and China from 2018 to 2021. This growth was further amplified by the Covid-19 pandemic.

Many of the key players in the e-grocery business are omnichannel grocers having their roots in classical stationary grocery retailing, see \citep{Wollenburg2018} for a review of the transition from ``brick-and-mortar'' to a ``brick-and-clicks'' grocery retailing and the implications for the underlying logistics networks. When grocers started their e-grocery business, they predominantly resorted to so-called in-store picking where the online orders are picked in existing brick-and-mortar stores located close to the customer from which they are then delivered to the customer. While many companies still employ this picking strategy in rural regions, it is not well suited to deal with the increasing volume of e-grocery purchases in large cities or metropolitan regions. As a consequence, aiming at increasing the efficiency of the picking process, which, according to our business partner, accounts for a substantial share the total warehousing costs, major e-grocery retailers started to establish dedicated warehouses, so-called fulfilment centres or ``dark stores'', that are only used for picking e-grocery orders. While \citet{hubner2023decision} develop a model for shelf space management in the light of real-world replenishment processes in the context of grocery retailing, research on e-grocery warehousing is still limited.

The process of warehousing is a key challenge for almost all retailers \citep{gu2007research}, and a suitable configuration of the warehouse depends on the assortment of the retailer under consideration, characteristics of stock keeping units (SKUs), as well as customer expectations, such as very high service level targets of 97-99\% \citep{Ulrich2018} and short delivery times, with some retailers even starting to offer same-day delivery. In e-grocery, most retailers offer an assortment of about 12,000 to 15,000 SKUs, some of which have special storage requirements such as the need for refrigeration. At the same time, an average order includes about 30 to 40 different products (order lines). While the need for short delivery times also arises in classical (non-grocery) e-commerce, in classical e-commerce, each customer order comprises only a few lines -- for example, according to \citet{Boysen2019}, each order at Amazon Germany comprises 1.6 lines. This difference in lines per order has a huge impact on the design and operation of a warehouse which is reflected by the fact that the recent review on warehousing for e-commerce by \citet{Boysen2019} explicitly excludes the e-grocery business.

In this paper, we deal with scientific decision support for a storage assignment problem arising in a fulfilment centre of a major European e-grocery retailer. While the retailer operates various fulfilment centres with different designs and degrees of automation, the one most recently established can be characterised as a hybrid or parallel warehouse in which a part of the assortment is allocated to a traditional picker-to-parts area. At the same time, the other SKUs are allocated to a partially automated area, where boxes sequentially move between stations within a so-called \textit{picking loop}\footnote{In the literature, this type of picking system is referred to as 'progressive zoning', see e.g.\ the review on warehouse order picking by \citet{deKoster2007}.}. At each station, a picker pulls the SKUs for a given customer from a shelve and places them into this box. In more detail, this configuration can be classified as a pick-and-pass system \citep{jane2000storage,pan2009study}.

For this fulfilment centre, we consider an integrated storage assignment problem involving three hierarchically related decisions: The first decision is to determine the subset of SKUs to be handled in the picking loop, the second is to determine the station of the picking loop to which a SKU is assigned, and the third one is to determine the shelf within the corresponding station. With the objective of establishing a high picking efficiency, the main goals associated with these decisions are to allocate SKUs with a high picking frequency to the loop, to keep the workload among the stations balanced, and to allocate high-demand SKUs close to the pickers in the stations. In addition, as it is typical for the retail business (see e.g. \citealp{trindade2022ramping}), the storage assignment needs to respect aspects such as the need for keeping space between SKUs located next to each other and precedence order constraints to ensure that heavy SKUs do not damage fragile ones in an order box.

As observed by \citet{Boysen2019} and many other authors, demand variation is one of the key challenges to be considered in almost every high-performance retail warehouse. If the demand for SKUs changes due to seasonal impacts or by long-term trends, maintaining a high level of picking efficiency typically requires to adapt the storage assignment by rearranging the storage locations of the SKUs. In case of short-term demand variations such as day-of-week dependent demand for certain SKUs, however, rearranging SKUs is often not possible or not meaningful -- as an example, this is the case in the e-grocery fulfilment centre considered in this paper. In order to mitigate the negative impact induced by such short-term demand variations, we propose to aim at determining a \textit{variation-aware storage assignment}, that is, a storage assignment that performs well for a number of demand scenarios, in particular for day-of-week-dependent SKU demands. The contributions of this paper can be stated as follows:

First, while there is a considerable amount of work on warehousing in non-grocery e-commerce and for brick-and-mortar retailing, see e.g.\ the recent reviews by \citet{Boysen2019} and \citet{Boysen2021}, our paper is among the first ones dealing with scientific support for tactical decision-making in e-grocery fulfilment centres. 
Second, we address a new three-level storage assignment problem occurring in a hybrid warehouse with a pick-and-pass system for fast order picking. For this problem, we propose and solve an integrated MILP formulation. In a set of experiments with real-world data provided by a leading European e-grocery retailer, we show that solving this integrated model is clearly superior to a standard sequential approach in which the selection of SKUs to be included in the fast-pick area is taken before taking zone/station assignment decisions.
Third, in order to cope with day-of-week dependent demand fluctuations, we propose to solve an augmented MILP model that explicitly aims at finding a storage assignment for the pick-and-pass system that performs well for each day of the week. Again using real-world data, we show that while a solution determined based on average demands leads to substantially imbalanced station workloads on certain days, the variation-aware solution is balanced on each day of week, almost without compromising the storage assignment objective.

The remainder of the paper is structured as follows: In the next section, we introduce the business case and the real-world data set considered in our study. Section 3 provides a review of related literature, followed by Section 4 covering our integrated storage assignment model and computational experiments. In Section 5, we develop a model extension taking into account varying demand patterns with respect to days of week of the SKUs. Finally, we sum up our major findings in the conclusion.

\section{Description of the logistic processes and available data}
\label{sec:setting}
The e-grocery retailer analysed in this paper mainly operates with a two-stage distribution process. In the first step, SKUs are supplied from national distribution warehouses to local fulfilment centres. These supplies typically take place at each working day between Monday and Saturday. When a supply arrives at a fulfilment centre, all SKUs are stored in their allocated shelves. In the second step, customer purchases are served by these fulfilment centres depending on the location of the customer. In the following, we describe the logistic processes within the fulfilment centre and the SKU data available in detail. Finally, we present historical picking data from the retailer with a focus on recurring patterns depending on the day of week.

\subsection{The process of order picking}
In most fulfilment centres, the retailer operates with a traditional picker-to-parts system. To improve operational efficiency, allow for shorter operation times, and increase the number of purchases served within one day, the retailer introduced a higher level of automation within certain fulfilment centres. While a fully automated picking process is cost-intensive, in this paper, we consider the case of a partially automated picking loop within a hybrid warehousing system established within one of the retailer's fulfilment centres. This hybrid system consists of two storage areas, (1) a partially automated picking loop and (2) a traditional picker-to-parts area. While the operational efficiency is higher within the picking loop, its available storage space is limited. As a result, the retailer has to decide which SKUs should be included in the picking loop and which should remain in the picker-to-parts area. Considering that an average customer order includes about 30 different SKUs, in general no order can be completed by only one of the storing areas. Instead, assembling all SKUs for a single purchase usually requires two independent picking processes in both areas. While there is comprehensive literature on traditional picker-to-parts area (see e.g.\ \citealp{caron1998routing}, and \citealp{franzke2017investigation}), the majority of picks in the warehouse of the business partner under consideration is performed within the picking loop. At the same time, the optimisation of the picking loop is more complex due to the existence of different stations. Therefore, in this paper, we focus on optimising the picking process within the more crucial pick-and-pass area.

\begin{figure}[htb!]
    \centering
    \includegraphics[width=0.6\textwidth]{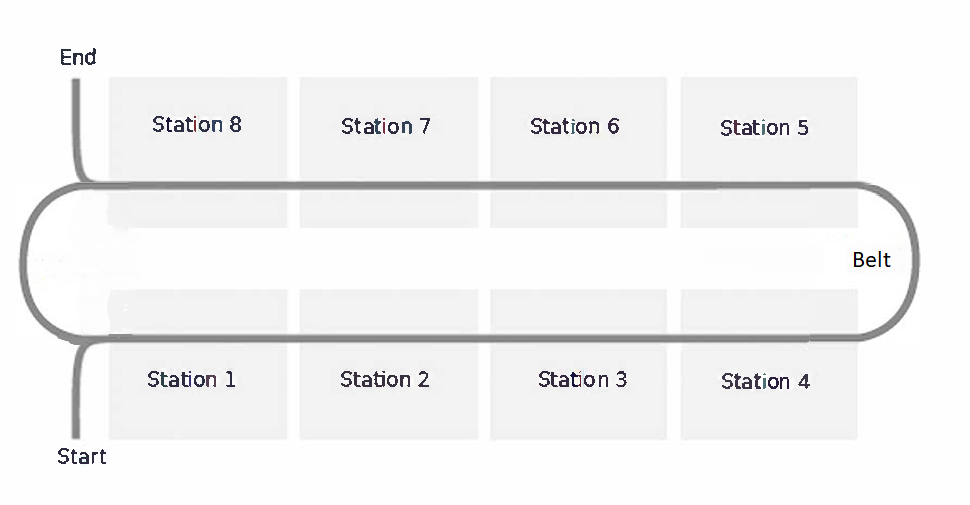}
    \caption{Representation of the picking loop.}
    \label{fig:picking_loop}
\end{figure}

The picking loop consists of eight picking stations, with boxes sequentially visiting the stations. Each box corresponds to one customer purchase, while at each station the picker pulls the SKUs for this purchase from the shelves and places them into the box. Once all SKUs corresponding to a specific customer purchase are placed into the box, it leaves the loop and the purchase is loaded into a vehicle for delivery. Figure \ref{fig:picking_loop} provides a schematic sketch of the picking loop.

\begin{figure}[htb!]
    \centering
    \includegraphics[width=0.4\textwidth]{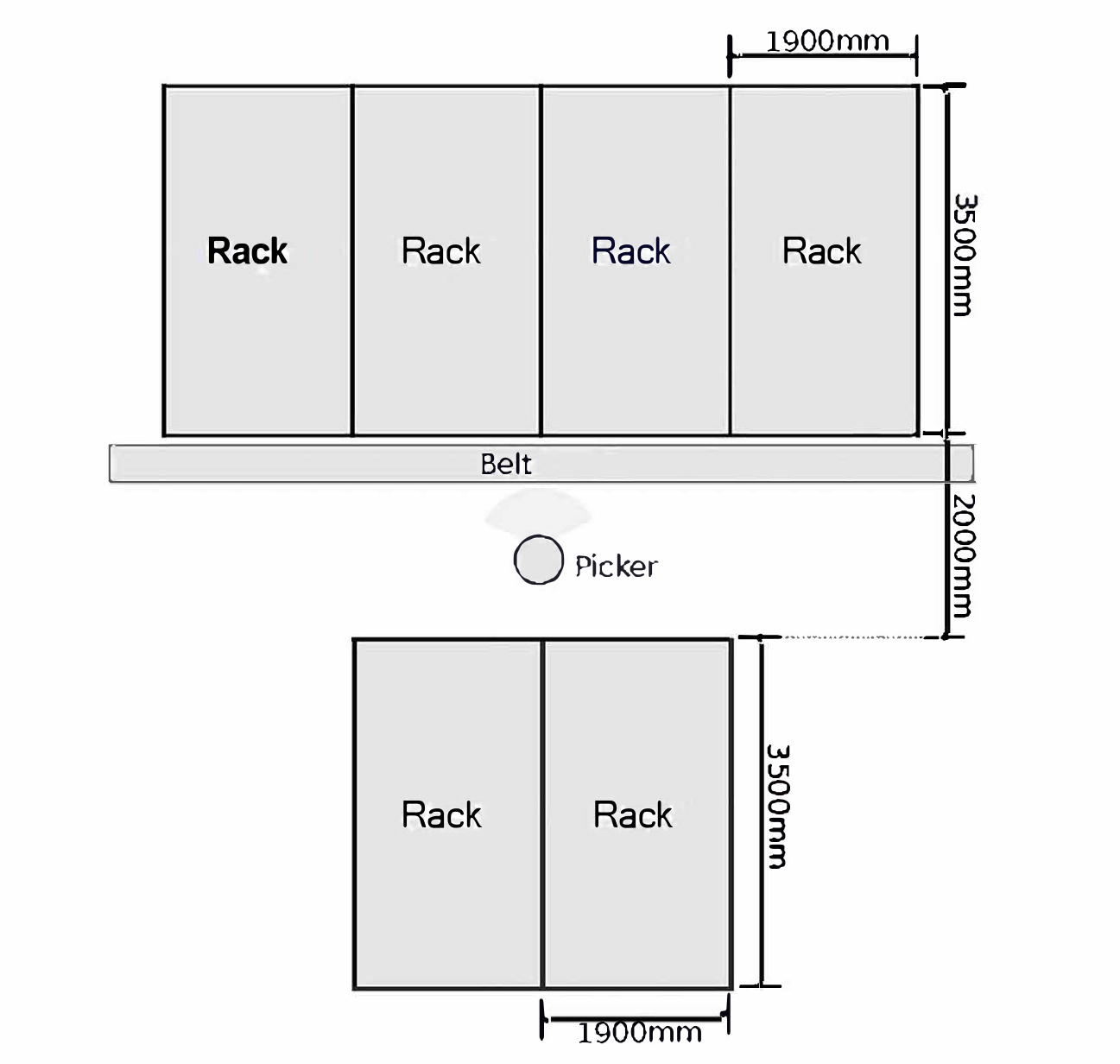}
    \caption{Representation of the structure of picking stations.}
    \label{fig:storages}
\end{figure}

Figure~\ref{fig:storages} illustrates the structure of an exemplary station within the picking loop. Each station consists of six racks, four in front of the picker, two outer and two inner, as well as two racks in the back of the picker. The racks in front of the picker contain four shelves with a height of 250 mm each (type 1), while the four shelves within the racks in the back of the picker have a height of 450 mm (type 2). In total, there are 192 shelves within the whole picking loop, 128 of type 1 and 64 of type 2. The structure of shelves is represented in Figure~\ref{fig:shelves} in the Appendix. Note that some SKUs can be allocated to type 2 shelves only due to their individual height.

To avoid congestion within the picking loop and idle times at some stations, the retailer aims at balancing the processing time and the corresponding workload for pickers between stations. For a given order, the time spent at a station depends on the number of picks, and on the shelf locations of the picked SKUs within the racks. While it is easy to pull SKUs out of a shelf at face level and in front of the picker, the picking process is more time-intensive for SKUs located in the top or bottom shelves of a rack as well as in shelves in the racks in the back of the picker. Therefore, in addition to the decision on allocating SKUs to the picking loop in general, the retailer needs to assign each SKU to a station and to a shelf with respect to the goals mentioned before. In contrast to brick-and-mortar retailing, where the allocation of SKUs to shelves also depends on marketing aspects, the setting of online retailing provides the company with the flexibility to decide on the placement of SKUs based on efficiency only. However, the retailer needs to take into account some additional constraints implicitly considered by customers in brick-and-mortar retailing, such as that large and heavy SKUs need to be placed into the box first in order to mitigate the risk of damaging fragile items. In order to avoid that the economic burden of picking heavy items in the first stations is allocated to a few pickers, the pickers rotate between stations during the day.

\subsection{SKU data}

The data set provided by the e-grocery retailer covers 4,693 different SKUs in total. It gives information on the measurements of each SKU, determining whether the SKU can be allocated to shelves of type 2 only or also to shelves of type 1. In addition, each SKU is associated with a precedence order rank taking one of the values 1, 2, and 3, where rank 1 corresponds to heavy items which need to be allocated to an early station and rank 3 is used for fragile items. All other SKUs are associated with rank 2. Furthermore, each SKU has a target stock depending on expected customer demand. This target, along with the size of the SKU, determines the space in the shelf to be allocated to the SKU. In addition, the shelf allocation needs to consider handling-related aspects, such as the need to reserve space for a separator if two different SKUs are placed next to each other in a single shelf. As mentioned above, the available space within the loop is not sufficient to store all SKUs in the assortment of the retailer. The decision on which SKUs are included into the picking loop are based on an importance score for each SKU.

\begin{figure}[ht]
    \centering
    \subfigure[Log importance score]{\includegraphics[width=0.42\textwidth]{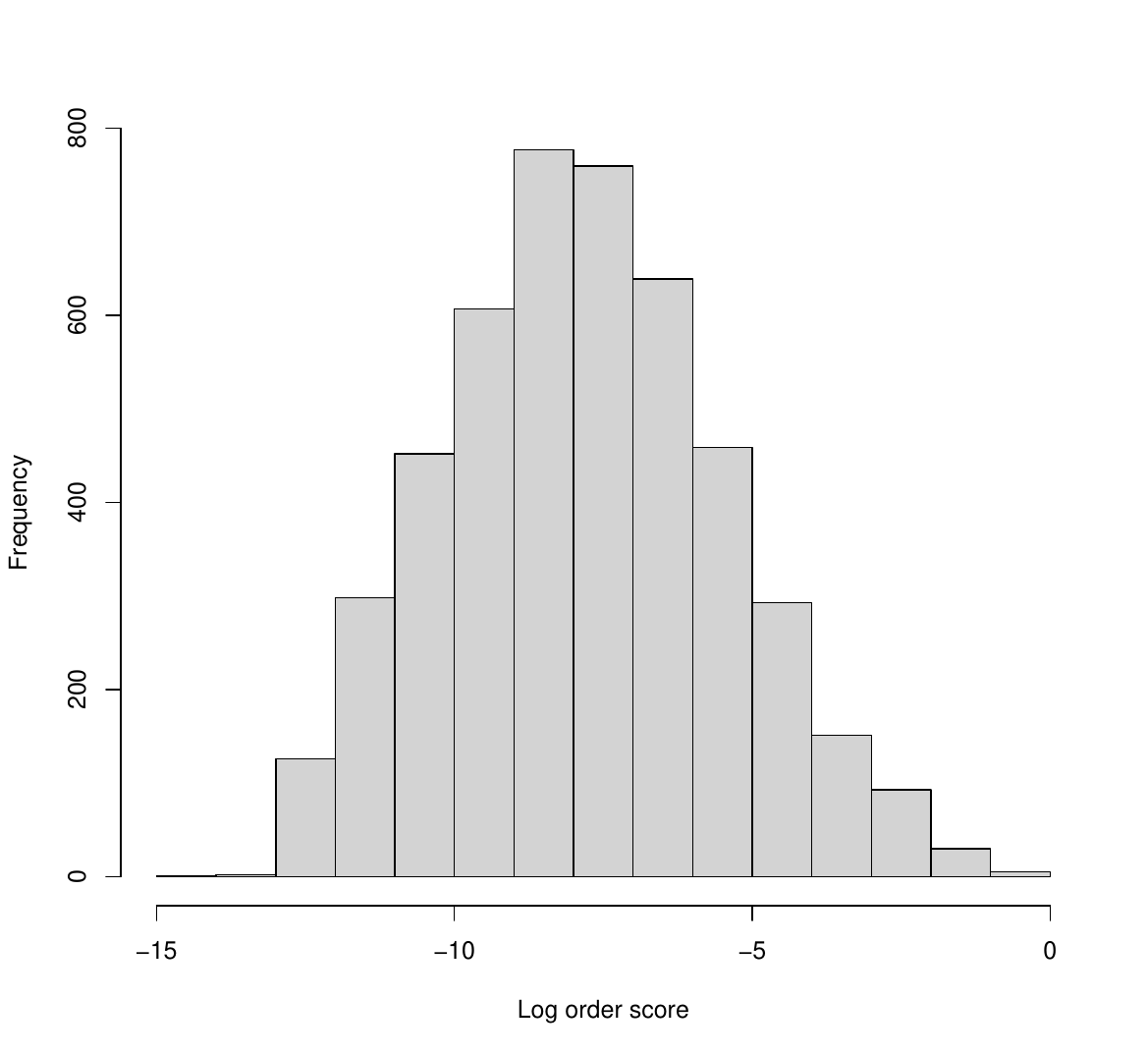}}
    \subfigure[Log number of picks]{\includegraphics[width=0.42\textwidth]{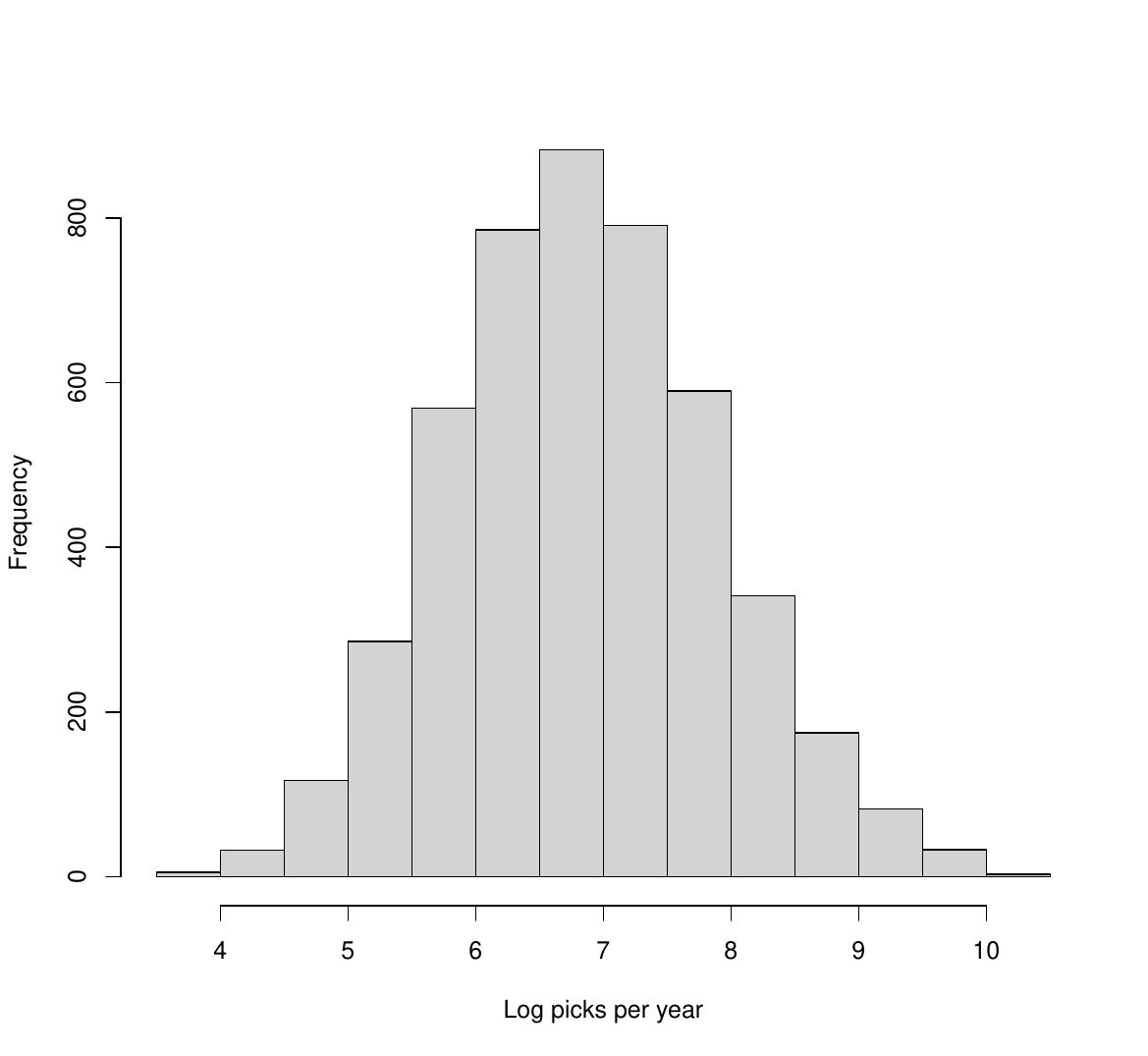}}
    \caption{Histograms of the log importance scores and the log number of picks for the SKUs in the assortment of the retailer appropriate for the picking loop.}
    \label{fig:order_scores_picks}
\end{figure}

This importance score, which is defined by the retailer, takes values between 0 and 1, where a higher value corresponds to higher importance of including this SKU into the picking loop. While this score is specific for the retailer under consideration, it allows for a generalisation of our model to other companies with their own importance measures. Here, the correlation between the importance score and the number of picks for the specific SKU (correlation coefficient of 0.718) should be noted. However, it also depends on additional factors such as the number of units per order and the volume of the SKU. As the distribution of the importance score has a strong positive skewness, we illustrate the frequency of the logarithm of importance scores for all SKUs in Figure \ref{fig:order_scores_picks} (a). The log importance score is approximately symmetric around its mean of -7.81 with standard deviation 2.34. This implies that there is only a small number of SKUs having an importance score exceeding 0.1, while the score is fairly small and nearly equal for a large majority of SKUs. Due to the positive skewness of the total number of picks for the SKUs in the assortment of the retailer appropriate to be included in the picking loop, we again show a histogram of the logarithm of the number of picks per SKU in Figure~\ref{fig:order_scores_picks} (b). This distribution is again roughly symmetric with a mean log number of picks of about 6.84. For more than 80\% of the SKUs, the average number of units per order line is at most 2 (mean 1.70). This confirms prior statements of \citet{Boysen2021} on the characteristics of e-grocery purchases. For some SKUs, however, the average number of units per order line is larger with up to 13.92 units (for details see the boxplot in Figure~\ref{fig:units_pick} in the Appendix). The target stock for SKUs again varies across the assortment. More than 90\% of the SKUs have a target stock of less than 20 units with an average of 9.10 units, implying some flexibility in the assignment due to the limited space needed to assign single SKUs. However, the 1\% with the highest target stock have an average of 100.35 and a maximum of 252 units. In addition, the measures of 17.4\% of the SKU require an allocation to type 2 shelves only.

\subsection{Historical picking data}
\label{sec:pikcing_data}
\begin{figure}[ht]
    \centering
    \includegraphics[width=0.5\textwidth]{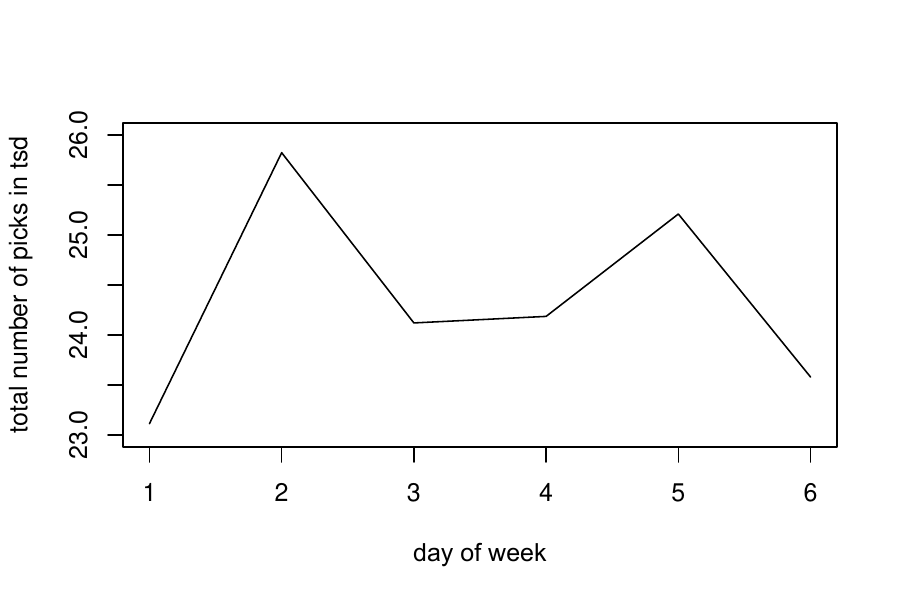}
    \caption{Total number of picks in thousands in the assortment of the retailer appropriate for the picking loop depending on the day of week (1 equals Monday, 6 Saturday).}
    \label{fig:average_picks}
\end{figure}

In addition to the characteristics of SKUs introduced above, the data set of the e-grocery retailer covers historical picking data and provides us with information on the average number of picks per month for a specific day of week for the SKUs within the assortment of the retailer, which are appropriate for the picking loop, for the year 2020. The data set covers the \textit{ID} of the SKU, the \textit{day of week} (1 corresponds to Monday, 6 to Saturday), the \textit{month}, and the corresponding average number of \textit{picks} for this month and day of week.\footnote{Note that the data is anonymised by multiplying the number of picks by the same constant for each entry, such that the relative relationship remains unchanged.} Positive values on Sundays correspond to picks at early Sunday morning if the purchases on the preceding Saturday could not be fully accomplished until midnight. Given 4,693 SKUs with data for 12 months each, we find 2,348 out of these 56,316 combinations with 0 picks for all days of the week, while in more than 30\% of the combinations we have 0 picks for at least one day of the week. This suggests the presence of variation in demand across days of week and months. Considering the average number of picks per day of week given in Figure~\ref{fig:average_picks}, we find peaks on Tuesday, driven by demand by business companies, as well as on Friday, where leisure goods are mainly demanded. These findings are supported by Table~\ref{tab:maxes} displaying the relative distribution of picks per day of week and additionally covering the number of SKUs where the highest demand is observed on this specific day of week. Again, we find the highest values for Tuesday, followed by Friday.

\begin{table}[ht]
    \centering
    \scalebox{0.65}{
    \begin{tabular}{c|cccccc}
        day of week & Monday & Tuesday & Wednesday & Thursday & Friday & Saturday \\
        \hline
        average relative number & 15.6\% & 17.3\% & 16.8\% & 16.9\% & 17.1\% & 16.3\% \\
        of picks on this day of week & & & & & & \\
        \hline
        number of SKUs with highest & 556 & 1127 & 702 & 746 & 919 & 643 \\
        demand on this day of week & & & & & & \\
        \hline
        number of SKUs with exceptional & 55 & 45 & 28 & 35 & 49 & 35 \\
        high demand on this day of week & & & & & & \\
    \end{tabular}}
    \caption{Average relative number of picks per day of week, number of SKUs where the highest demand is given on this day of week and number of SKUs with exceptional high demand on a specific day of week.}
    \label{tab:maxes}
\end{table}

\begin{figure}[ht]
    \centering
    \subfigure[Boxplots]{\includegraphics[width=0.45\textwidth]{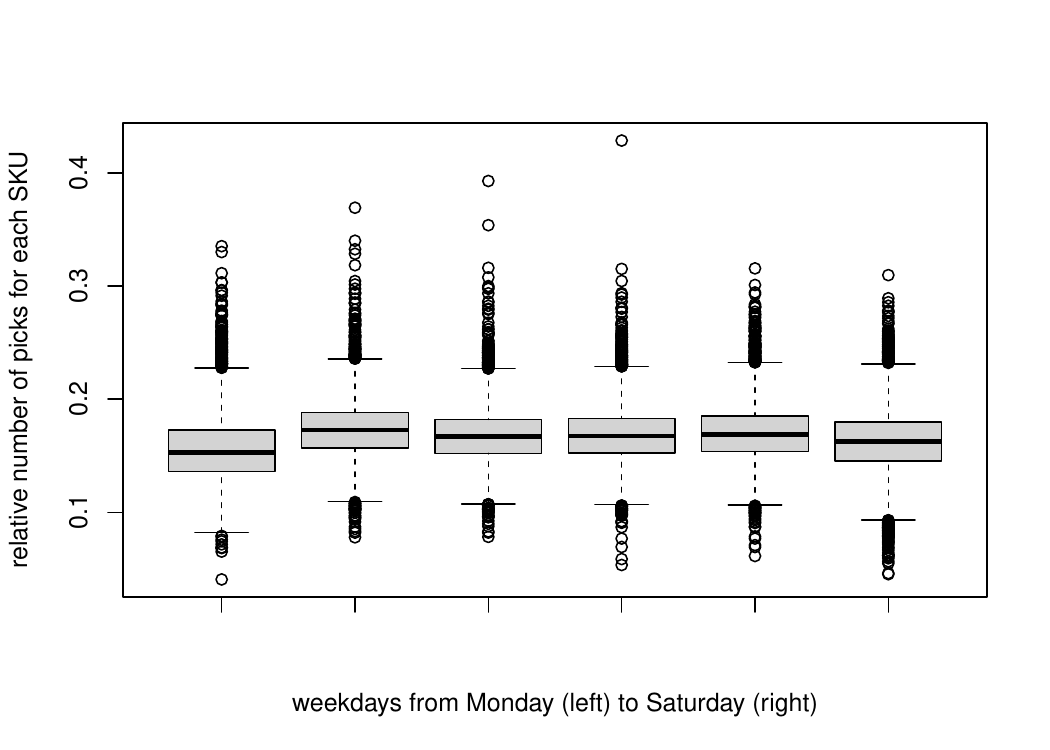}}
    \subfigure[Detailed description]{\includegraphics[width=0.45\textwidth]{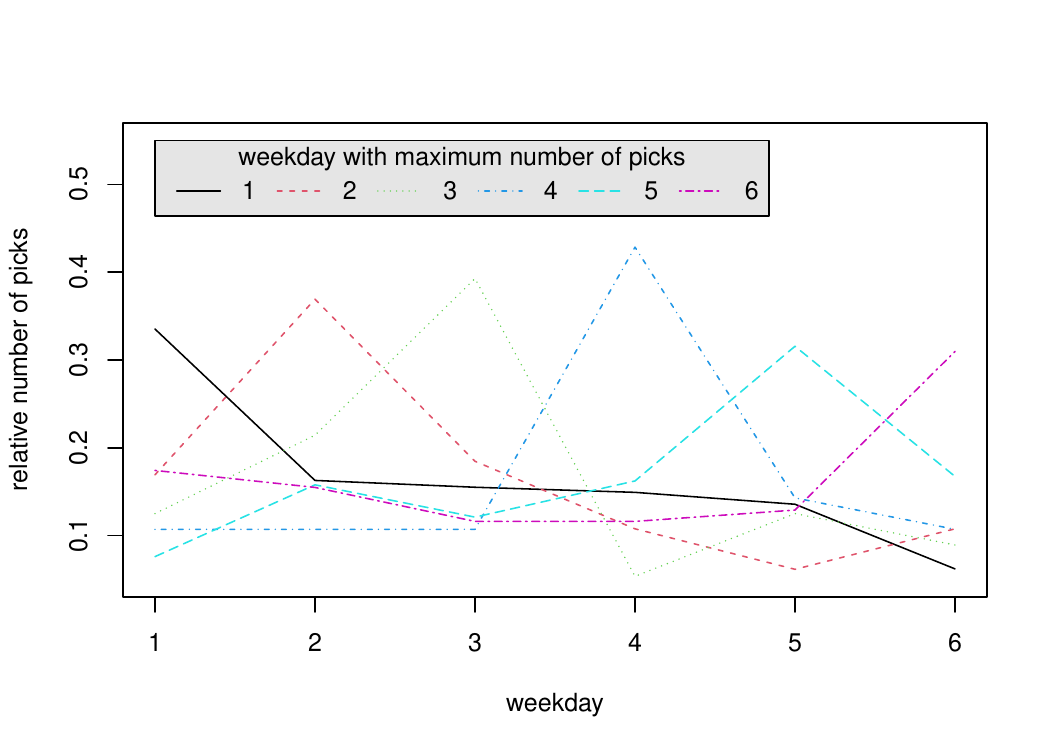}}
    \caption{Boxplots of the relative number of picks and detailed description of SKUs corresponding to those with the highest relative number of picks at a specific day of the week in the assortment of the retailer appropriate for the picking loop.}
    \label{fig:avg}
\end{figure}

It should be noted that a constant proportional change in demand over all SKUs included in the picking loop would not affect the balancing between stations significantly. However, if there is high demand for multiple SKUs in the same station compared to other stations on a specific day of week, this would incur congestion at this station and, therefore, affect the operational efficiency of the retailer. As an example, SKUs with high demand at the beginning of a week should be matched with those predominantly demanded right before the weekend to balance the workload across stations. Table~\ref{tab:maxes} confirms the results of Figure~\ref{fig:average_picks} and suggests that SKUs can be categorized into two main groups, one with highest demand at the beginning of the week and one with highest demand at the end of the week, while the relative number of picks per day of week over all SKUs fluctuates between 15.6\% and 17.3\%. Therefore, we define an exceptional high demand at a certain day of week if more than 25\% of the picks per week for the specific SKU are accomplished on this day of week. Again, we find high values for Tuesday and Friday, but also for Monday. Figure~\ref{fig:avg} (a) gives boxplots on the relative number of picks for each SKU on a specific day of week. While there is a higher variance for Monday and Saturday, for the other days we find 50\% of the SKUs to have a relative amount of weekly picks between 15\% and 19\%, whereas on Wednesday there is one SKU with a relative number of picks exceeding 40\%. In more detail, Figure~\ref{fig:avg} (b) covers the six SKUs corresponding to those having the maximum relative amount of weekly picks at a specific day of week.\footnote{We exclude SKUs for which data is not available for single days of week.} We display their relative distribution of picks for each day of week, indicating that there is strong variation in the number of picks for these SKUs across different days of the week.

\section{Related work}
\label{sec:related_work}
As discussed by \citet{Hubner2019}, establishing efficiency in distribution logistics is one of the most challenging and success-critical tasks for e-grocery retailers. For recent overviews on the specific challenges faced by omni-channel and e-grocery retailers, and the resulting implications on the design and operation of their logistics networks, see e.g.\ \citep{Wollenburg2018,Hubner2019, RodriguezGarcia2021}. Given these challenges and the increasing relevance of e-grocery retailing, it comes to no surprise that the Operations Research community recently started to develop optimisation-based approaches for supporting decision-making in distribution logistics.

\subsection*{Decision support for e-grocery warehousing}
To give a few examples for a fulfilment process based on in-store-picking, \citet{Vazquez-Noguerol2022} propose an optimisation model that allocates customer orders to stores where these orders are picked, and schedules both, the picking as well as the delivery of the order to customers. \citet{Dethlefs2022} consider a setting where the e-grocer operates with both in-store picking and picking in distribution centres, proposing an approach that integrates the assignment of orders to stores or distribution centres and the scheduling and routing of deliveries. A recent overview on the new role of brick-and-mortar stores in omnichannel retailing is given by \citet{Hubner2022}. However, we are not aware of any paper dealing with optimisation-based decision support for tactical problems such as storage assignment in dedicated e-grocery fulfilment centres. This gap in the research is particularly relevant since, as detailed by \citet{Hubner2019}, e-grocery fulfilment is substantially different from fulfilment for non-grocery e-commerce and from warehousing in distribution centres supplying grocery stores. These differences are also discussed in the recent reviews on warehousing for e-commerce by \citet{Boysen2019} and for brick-and-mortar retailing by \citet{Boysen2021}, both of which explicitly exclude the case of e-grocery fulfilment warehouses. To illustrate the differences between e-grocery and other e-commerce characteristics and their impact on warehousing, let us consider the usefulness of a fast-pick area based on a pick-and-pass system: Given the fact that in classical e-commerce settings, each customer purchase only involves a few number of order lines, \citet{Boysen2019} consider the use of a pick-and-pass system as inappropriate for e-commerce warehouses. An e-grocery order, however, typically involves dozens of order lines which completely changes this assessment: In fact, a leading European e-grocery retailer decided to use a variant of such a system as part of the fulfilment centre we consider in this paper.

\subsection*{Storage assigment}
Using this warehouse as underlying business case, we propose an integrated approach for taking the three main storage assignment decisions identified in the review paper by \citet{deKoster2007}: allocating SKUs to a specific area of the warehouse (in our case, either to a standard picker-to-parts based area in the warehouse or to the fast-picking area), assigning SKUs to zones of a given area (in our case, stations in the pick-and-pass based picking loop), and the assignment of SKUs to shelves within a given zone (in our case, the picking station). While the exact problem considered here, which, as described above, also involves specific aspects such as order constraints, has not been discussed in the literature so far, we are also not aware of any work that addresses all three storage assignment decisions in an integrated way in other warehousing settings. Next, we will briefly review existing research dealing with taking storage assignment decisions, with an emphasis on pick-and-pass systems considered in this paper.

The highest-level storage assignment decision considered in this paper is which SKUs to assign to the highly efficient picking loop system and which to assign to the picker-to-parts area of the warehouse. A similar decision arises in the so-called \textit{forward reserve allocation problem} (see e.g. \citealp{Hackman1990} and \citealp{Walter2013}) which consists in selecting the SKUs to be allocated to the fast-picking area along with the number of units to be allocated of each selected SKU. In that problem, it is assumed that the fast-picking area is re-filled from the reserve area, whereas in our setting each SKU is assigned to a single location in one of the two parts of the warehouse. In contrast to the integrated problem considered in our paper, the forward-reserve allocation problem does not consider the assignment of SKUs to storage locations but only aims at ensuring that the SKUs allocated to the fast-picking area can be assigned to the shelves.

The next decision to be considered in our problem is the the assignment of SKUs to zones or stations in the picking loop. As discussed in the review by \citet{deKoster2007}, when it comes to assigning SKUs to zones or stations in a pick-and-pass system (also referred to as ``progressive zoning''), the most important goal is to balance the workload among the zones. In fact, workload balancing is either part of the objective or the constraint set in most of the works dealing with the optimisation of zone assignment decisions, see e.g. \citep{jane2000storage,Jewkes2004,deKoster2008, hong2016order}. Actually, the positive effect of workload balancing on the performance of pick-and-pass systems was verified in several studies using simulation \citep{jane2000storage, deKoster2008, Pan2015} and approximate models based on queuing theory \citep{Yu2008, Pan2015, vanderGaast2020}. With the exception of \citep{jane2000storage}, the majority of the papers dealing with storage assignment in pick-and-pass systems combine the allocation to zones with storage location assignment in shelves. In general, the main goal of this shelf assignment is to determine the shelf locations in a way that SKUs with a high picking frequency have a short picking time. Instead of focusing on picking efficiency, \citet{Otto2017} propose to focus on ergonomic aspects, and, similar to \citep{Jewkes2004}, consider an order line system in which the configuration of the zones in terms of zone borders/allocation of rack columns is part of the decision problem (which is not the case in our setting).

\subsection*{Dealing with demand variation}
Irrespective of the type of picking system, the majority of the articles dealing with storage assignment assume a given demand scenario for which the assignment is optimised. In practice, however, demand is subject to variation: In case of the e-grocery retailer considered in this paper, for example, there are multiple demand variation patterns depending on the day of week and on the season, and there are long-term trends leading to structural changes in customer buying behaviour (e.g.\ increased demand for vegan and vegetarian products). Clearly, as noted by \citet{Pazour2015}, a storage assignment which is optimal for a given demand scenario may be suboptimal for another scenario. As a consequence, there are several papers such as those by \citet{Christofides1973}, \citet{Chen2011}, or \citet{Pazour2015} dealing with rearranging the storage location configuration of warehouses. While such adaptation of the storage assignment is useful in case of long-term demand fluctuations, when it comes to short-term demand variations, such an adaptation is typically not feasible or meaningful from an economical perspective.

Research on variation-aware storage assignment is relatively scarce and mostly considers warehouse designs that substantially differ from the setting of the business partner considered in this paper: For the case of a unit-load warehouse, \citet{ang2012} deal with finding a storage allocation policy in presence of varying demands. They show that their policy obtained using robust optimisation significantly outperforms variation-agnostic policies from the literature when it comes to expected performance. For a warehouse storing pallets from a car parts manufacturer, \citet{Kofler2015} discuss a robust storage reallocation strategy, that is, one that is robust to small demand variations, thereby reducing the need for storage reallocations. When it comes to pick-and-pass warehouses, we are not aware of any article dealing with finding a storage assignment that performs well in presence of short-term demand variations.

\section{Integrated three-level storage assignment}
\label{sec:basic-model}

In this section, we first propose a MILP model formulation that simultaneously considers the three decisions outlined above: Selection of SKUs to be included in the picking loop, assignment of SKUs to picking stations, and assignment to the shelves in the station. In Subsection~\ref{sec:basic_comp}, we present the results from a series of experiments with this model using real-world data.

\subsection{Problem description and model formulation}
\label{sec:basic_model_form}
Our problem consists of three main decisions: First, given a set $V$ of SKUs and a hybrid warehouse consisting of a picker-to-parts area with relatively low picking efficiency and a picking loop with a high picking efficiency, we have to decide which SKU to allocate to the picking loop.\footnote{Note that among all SKUs in the assortment of the retailer, $V$ only comprises those that can be handled by the picking loop  -- as an example, $V$ does not include any SKUs that need to be refrigerated.} This allocation is based on a so-called importance score $s_v$ associated with each SKU $v$ which, as described in Section~\ref{sec:setting}, in our case study is provided by the retailer. If we consider only this first decision and aim at maximising the sum of the importance scores, the resulting problem is very similar to the so-called forward-reserve problem reviewed in Section~\ref{sec:related_work}. Note, however, that while in the forward-reserve problem it is assumed that the reserve area serves both as a picking area for the SKUs not assigned to the fast-picking zone and as reserve area from which the fast-picking system is restocked, in the problem considered here, each SKU is either assigned to the picking loop or to the picker-to-parts area.  Consequently, it is assumed that the (fixed) shelf space taken by an SKU $v$ (characterised by its height $h_v$ and the width $w_v$) is large enough to store all units of the SKU until the next re-supply of the SKU $v$. In this context, observe that $w_v$ is not the width of a unit of a SKU but the width of the target stock of $v$ if it is included in the picking loop; for the problem considered here, $w_v$ is considered as a given and fixed parameter.

For the SKUs to be allocated to the picking loop, the second decision to be considered is the assignment of the SKU $v$ to a station $k$ from the set $K$ of stations which we assume is ordered and indexed by integers ($K = \{1 ,..., |K|\}$). This station assignment, which can be viewed as a zone assignment in a pick-and-pass system, needs to consider two main aspects: First, the assignment has to respect a set of precedence order constraints. As described above, we assume that each SKU is associated with a precedence rank $o_v \in O $, with $O = \{1 ,..., |O|\}$. In order to avoid damaging of SKUs, the station assignment decision needs to make sure that for each pair of two SKUs $v$ and $v'$ with $o_v \leq o_{v'}$, $v$ is assigned to the same or an earlier station as $v'$. In addition to respecting these precedence order constraints, the workload among the stations in terms of number of picks should be balanced. Here, following the requirement of our business partner, we assume that there is a given threshold $\delta$ denoting the maximum permitted relative deviation of the workload $z_k$ of a station $k$ from the average value over all stations in terms of picking operations per day. This average workload can be calculated as $z = \frac{1}{|K|} \sum\limits_{k \in K} z_k$.

The third storage assignment decision is the assignment of SKUs to shelves within the stations in a way that SKUs with a high number of picks are stored in shelves that are fast and easy to reach by the picker. It is assumed that each SKU is assigned to a single shelve. We denote the set of shelves at a given station $k$ with $R_k$ and the set of all shelves in the picking loop with $R$, that is $R= \bigcup\limits_{k \in K} R_k$. Each shelf $r \in R$ has a height $h_r$ and a width $w_r$, and an SKU $v$ can only be assigned to a shelf $r$ if $h_v \leq h_r$ and $w_v \leq w_r$, giving rise to the definition of a set $R^v \subseteq R$ of shelves that can fit SKU $v$. Note that in addition to this, we can further restrict the set $R^v$ based on the precedence rank $o_v$: As an example, if all SKUs with rank $1$ fit into the first two stations, this implies that an SKU $v$ with $o_v =1$ cannot be assigned to a shelf in a station $k > 2$ without either leaving shelf space empty in the first two stations or violating the precedence constraints. If more than one SKU is stored in a shelf, there needs to be a minimum distance $g$ between each two SKUs stored next to each other. In addition, a shelf $r$ is associated with a distance $d_r$ from the picker at the corresponding station $k_r$. We assume that every pick is carried out separately, and, following the definition used by the management of the e-grocery retailer, we define the picking efficiency of a shelf $r$ as the inverse of the  distance $d_r$, that is, as $\frac{1}{d_r}$. Using this definition, the goal in this subproblem is to allocate SKUs with a large number of picks in shelves exhibiting a high picking efficiency. Accordingly, in this subproblem we maximise the average efficiency per pick.

Since the three storage assignment decisions described above are highly interdependent, we aim at considering them simultaneously in an integrated problem. Among these three decisions, only the first and the third one are associated with an objective function, namely maximising the sum of the importance scores of the SKUs selected for the picking loop and maximising the picking efficiency of the shelf allocation. In the integrated problem, we combine these two objectives in form of a convex combination by introducing weights $\alpha$ to be multiplied with the first part and $1-\alpha$ to be multiplied with the second part of the objective function.

Next, we propose a MILP formulation for the integrated problem. The main decision variables in this formulation are the binary variables $x_{v,r}$ taking value 1 if SKU $v$ is assigned to shelf $r$ and 0 otherwise. The values of these variables determine the values  of the second set of variables considered in the model, namely the variables $z_k$ denoting the workload in terms of total number of picks assigned to station $k$. Furthermore, we introduce the integer variable $y_o$ representing the last station (that is, the station with the highest index $k$) to which a SKU with precedence rank $o$ is assigned. Given these variables and the parameters introduced above, we are now ready to present the MILP formulation of the integrated problem:\footnote{For a table containing all notation of the model see Table~\ref{MILP_basic} in the Appendix.}

\begin{equation*}
        \max \frac{\alpha}{\gamma_1} \underbrace{\sum\limits_{v \in V}\sum\limits_{r \in R^v} s_v x_{v,r}}_{\substack{I}} + \frac{1-\alpha}{\gamma_2} \underbrace{ \sum\limits_{v \in V}\sum\limits_{r \in R^v} \frac{1}{d_r} p_v x_{v,r}}_{\substack{II}}
\end{equation*}

\noindent
\begin{align}
        \sum\limits_{r\in R^v} x_{v,r} & \leq 1 \,\,\, &&\forall \,\,\, v\in V \\
        k_r  x_{v,r} &\leq y_{o} \,\,\, &&\forall \,\,\, o \in O, v \in V^{o}, r \in R^{v} \\
        k_r  x_{v,r}  &\geq y_{o-1} \,\,\, &&\forall \,\,\, o  \in O  \setminus \{ 1 \}, v \in V^{o}, r \in R^{v} \\
        z_k &= \sum\limits_{v \in V}\sum\limits_{r \in R_k} p_v x_{v,r} \,\,\,  &&\forall \,\,\, k \in K \\
        z_k & \leq (1+\delta) \cdot \frac{1}{|K|} \sum\limits_{l \in K} z_l \,\,\,  &&\forall \,\,\, k \in K  \\
        z_k & \geq   (1- \delta) \cdot \frac{1}{|K|}  \sum\limits_{l \in K} z_l    \,\,\,  && \forall \,\,\, k \in K  \\
        w_r & \geq \sum\limits_{v \in V} (w_v + g) \cdot x_{v,r} - g \,\,\, &&\forall \,\,\, r\in R\\
        x_{v,r} & \in  \{0, 1\} \,\,\, &&\forall \,\,\, v \in V, r \in R^v  \\
        y_{o} & \in \{1,...,|K|\} \,\,\,& &\forall \,\,\, o \in O
\end{align}

The objective function consists of a weighted combination of the two parts mentioned above: Part I corresponds to the maximisation of the total importance score, while part II represents the maximisation of the average efficiency per pick. To bound both parts of the objective function and, therefore, also the total objective value in the interval [0,1], we normalise the objective function by dividing through $\gamma_1$ and $\gamma_2$, respectively. The parameter $\gamma_1$ corresponds to the objective value when optimising the total importance score (part I of the objective function) individually, $\gamma_2$ to the situation where the picking efficiency is optimised only. By adjusting the parameter $\alpha$, the relative importance of the two objectives can then be adjusted by the decision maker.

Constraint set $(1)$ ensures that each SKU is assigned to at most one shelf in the picking loop. Constraints $(2)$ and $(3)$ enforce the precedence order constraints: Constraint set $(2)$ imposes that $y_o$ is at least as big as the maximum station index $k_r$ of a shelf $r$ a SKU with order rank $o$ is assigned to, and $(3)$ ensures that all SKUs with a precedence rank $o$ other than 1 are assigned to a station $k \geq y_{o-1}$, that is, to a station corresponding to the last station containing a SKU with the next smaller rank or to a station later in the loop. The Constraints $(4)$ -- $(6)$ enforce balanced workload among the stations. Constraint set $(4)$ is used to determine the value of the auxiliary variables $z_k$ representing the total number of picking operations allocated to station $k$. Using this variable, Constraints $(5)$ and $(6)$ ensure that the workload allocated to each station respects the maximum permitted relative deviation from the average workload among all stations. Constraints $(7)$ ensure that the total width of the SKUs assigned to a shelf $r$ plus the required gaps between each pair of SKUs in a shelf does not exceed the width  $w_r$ of the shelf. Finally, the Constraints $(8)$ and $(9)$ enforce the domains of the variables $x_{v,r}$ and $y_o$.

\subsection{Computational experiments}
\label{sec:basic_comp}
In this section, we present the results from a number of experiments with the model presented above, using real-world data from the e-grocery retailer considered in this paper. In a first set of experiments, we explore the solution behaviour with respect to convergence of the duality gap, i.e.\ the relative difference between a solution found by the optimiser and an upper bound, over time. In addition, we discuss the impact of the weighting factor $\alpha$ on the values of the two parts of the objective function, given that we allow for a fixed relative deviation $\delta$ in the number of picks between stations. This allows us to derive an interval of reasonable values for the weighting factor $\alpha$. Furthermore, we consider the effect of the allowed deviation $\delta$ between stations on the structure of the solutions. Finally, we also compare our integrated three-level storage assignment approach to a sequential proceeding, i.e.\ solving the area allocation problem (akin to the forward reserve problem), the assignment to stations, and the assignment of the selected SKUs to shelves consecutively. All experiments were conducted with the Gurobi optimiser version 9.0.2 on a computer with 16 GB RAM and a AMD Ryzen\texttrademark~ 5 1600 3.2 GHz CPU.

\subsubsection*{Experiments on the runtime}
\begin{figure}[ht]
    \centering
    \subfigure[runtime vs. objective value]{\includegraphics[width=0.42\textwidth]{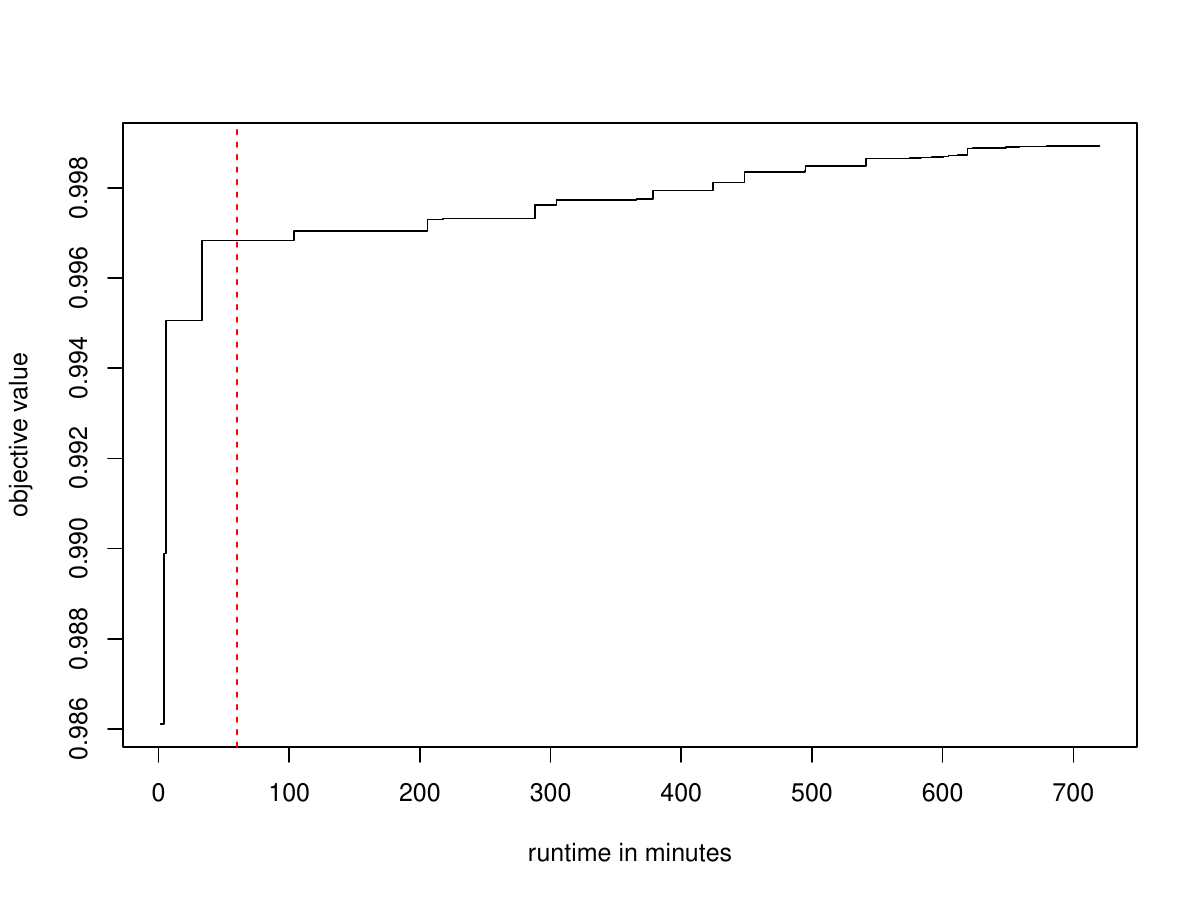}}
    \subfigure[runtime vs. gap]{\includegraphics[width=0.42\textwidth]{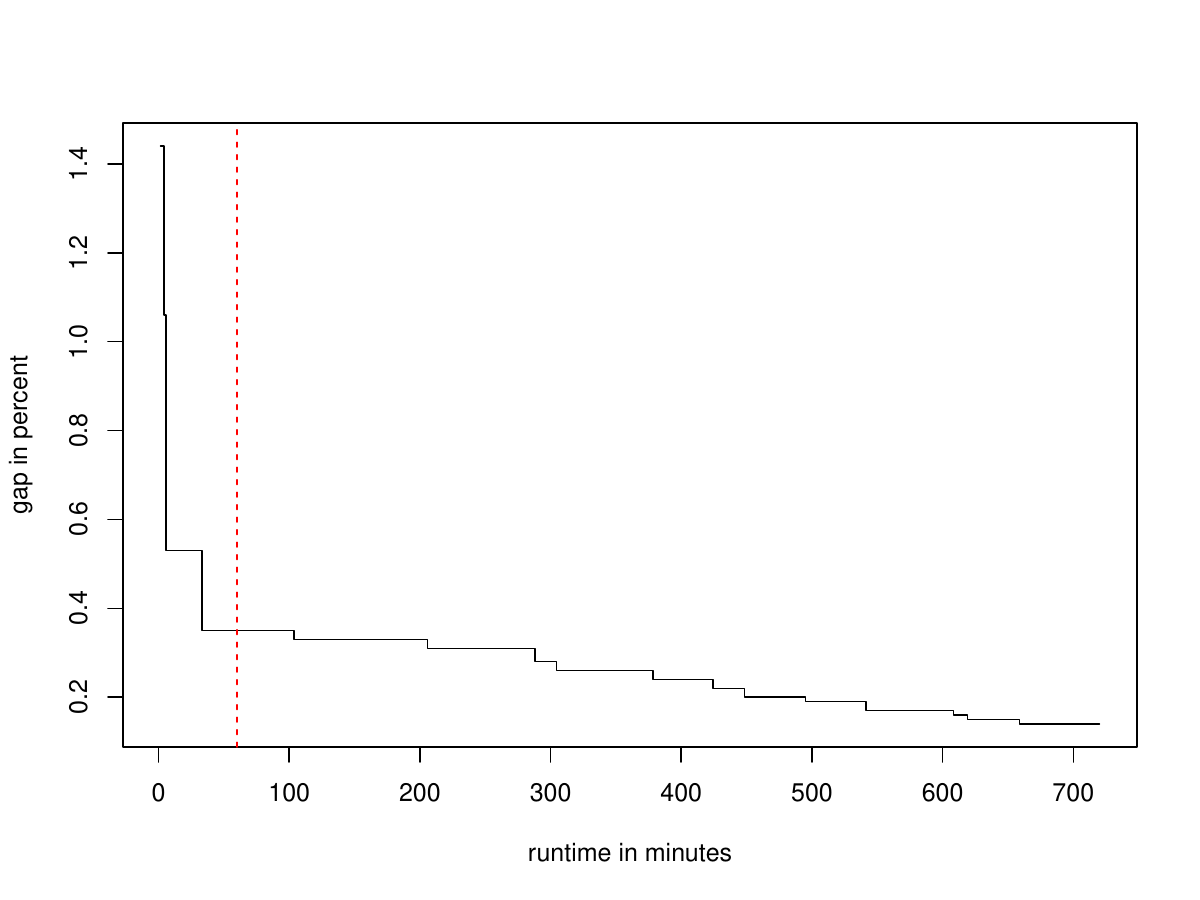}}
    \caption{Representation of the objective value and gap to the lower bound in percent depending on the runtime in minutes of up to 12 hours using $\alpha=0.5$ and $\delta=1\%$. The red dotted line corresponds to a runtime of 1 hour.}
    \label{fig:runtime}
\end{figure}

In a first analysis, we use an exemplary weighting factor $\alpha=0.5$ and allow for a deviation of picks between stations of $\delta=1\%$. Figure~\ref{fig:runtime} shows (a) the objective value and (b) the gap to the lower bound after a given runtime of up to 12 hours. We additionally emphasise the resulting values after a runtime of one hour by the red dotted lines. In this exemplary setting, after the intended runtime a gap of 0.35\% remains. As we cover a tactical problem of the retailer, that is not solved regularly, even longer runtimes could be allowed. However, our results show that the progress in further reduction of the gap is slow. For example, even after four additional hours of runtime the gap reduces by another 0.04 percentage points only.

\subsubsection*{Effect of the objective weight $\alpha$}
\begin{figure}[ht]
    \centering
    \subfigure[part I (importance score)]{\includegraphics[width=0.42\textwidth]{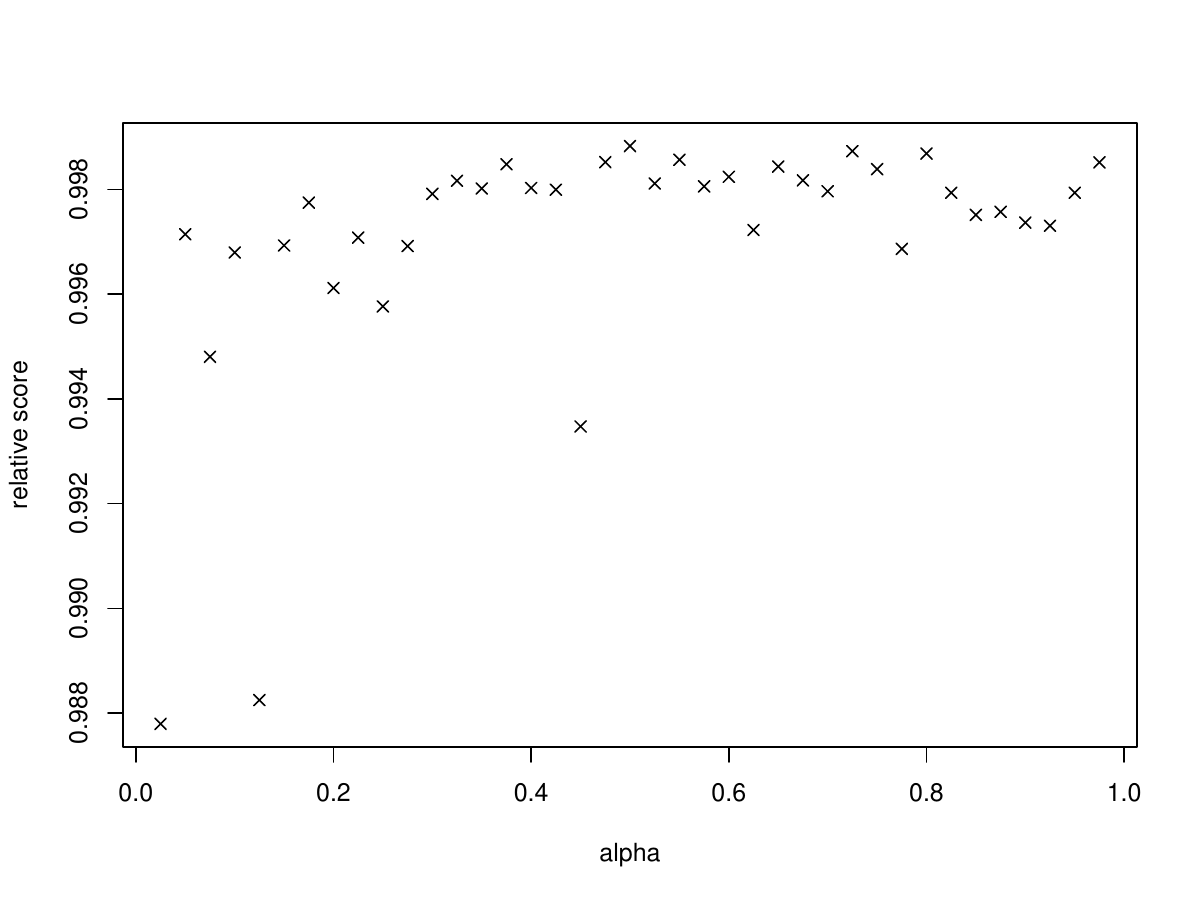}}
    \subfigure[part II (picking efficiency)]{\includegraphics[width=0.42\textwidth]{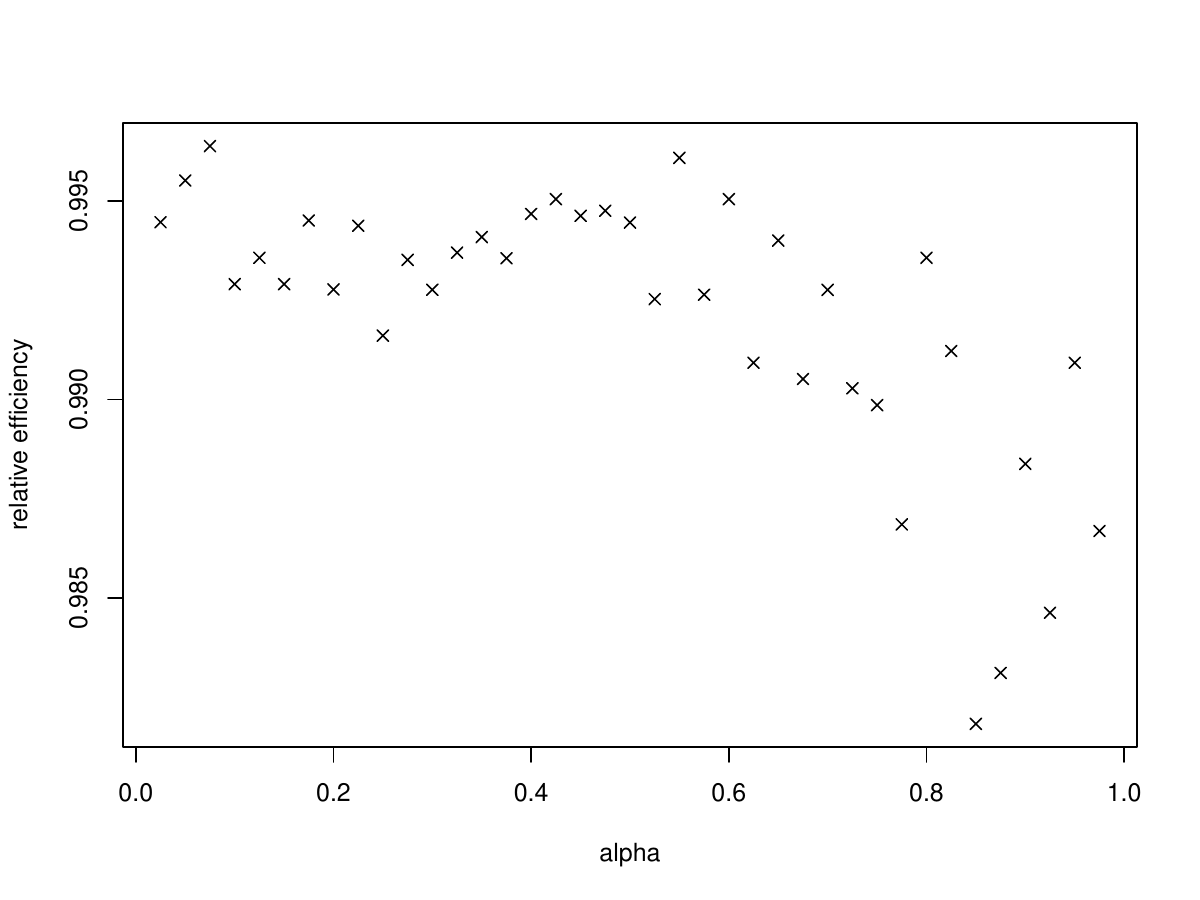}}
    \caption{Normalised values of part I (importance score) and part II (picking efficiency) of the objective function depending on the weighting factor $\alpha$ for $\delta = 1\%$.}
    \label{fig:beta_stern}
\end{figure}

As introduced above, the objective function consists of two parts, where the first (I) covers the sum of importance scores of SKUs allocated to the picking loop area and the second (II) controls for the picking efficiency in the picking loop. We consider the impact of different values for the weighting factor $\alpha$ on our results in a set of experiments. As this requires a comprehensive number of experiments, we terminate the optimisation when either a gap of 0.5\% or a predefined time limit of 30 minutes is reached to reduce computational effort, while we limit the relative deviation of picks between stations here to $\delta=1\%$. Figure~\ref{fig:beta_stern} gives an overview on the values for both parts of the (normalised) objective function depending on $\alpha$. For the reason of a more expressive visualisation, we exclude the resulting values for $\alpha=0$ (score 0.912; efficiency 0.997) and $\alpha=1$ (score 0.999; efficiency 0.381). The left part of the figure illustrates that the normalised sum of importance scores of the SKUs assigned to the picking loop takes the highest values for $\alpha \geq 0.35$. At the same time, part II of the objective function remains almost in the same interval for $\alpha \leq 0.6$ and decreases for large values of $\alpha$.\footnote{Note that we find an outlier for the score when using $\alpha=0.45$. However, this situation is driven by the limitation of the runtime here and wont occur in practice when allowing for a longer runtime.}

\begin{figure}[!htb]
    \centering
    \includegraphics[width = 0.7\textwidth]{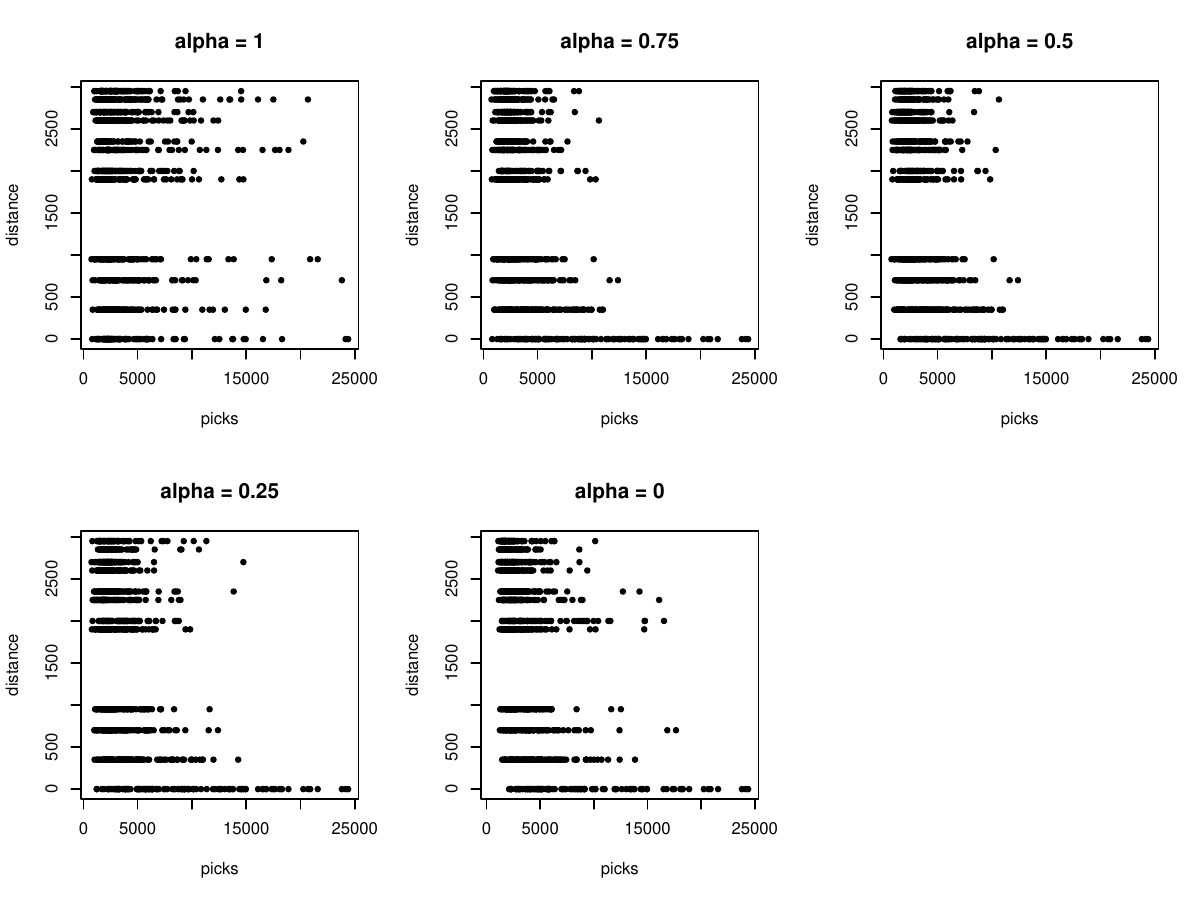}
    \caption{Allocation of SKUs and corresponding picks to shelves with given distance to the picker for different values of $\alpha$ and $\delta=1\%$. Note that the figure is limited to SKUs with a height of up to 250 mm.}
    \label{fig:distanz_vergleich}
\end{figure}

Figure~\ref{fig:distanz_vergleich} gives additional insights in the structure of the solutions by showing the number of picks for a given distance between the picker and the shelf for $\alpha \in \{0, 0.25, 0.5, 0.75, 1\}$. We exclude SKUs with a height exceeding 250 mm from these plots as they can be allocated to type II shelves only (see Figure~\ref{fig:distanz_vergleich_250plus} in the Appendix for the allocation to type II shelves). For $\alpha=1$ (i.e.\ a setting where the distance between the picker and the corresponding shelf does not affect the objective value), Figure~\ref{fig:distanz_vergleich} indicates a non-systematic pattern in the allocation of SKUs to shelves. At the same time, for $\alpha \leq 0.75$ SKUs with a high number of picks are allocated to shelves close to the picker, with results changing only slightly for smaller values of $\alpha$. However, there is still a small number of outliers in each figure. For example, given the allocation for $\alpha=0.5$, there are some SKUs with a high number of picks which are still allocated to shelves with a distance of 1900 mm and 2850 mm, respectively. These specific SKUs have a high width and, therefore, the model favours the allocation of more but smaller SKUs with a high number of picks over these SKUs to shelves close to the picker. For $\alpha=0$, i.e.\ a situation where we focus on picking efficiency only but exclude the importance score from the objective, we find a drop in the number of SKUs allocated to the shelve close to the picker, in particular for those SKUs having a small number of picks. At the same time, the total number of SKUs allocated to the picking loop increases by about 20\% and the total number of picks by 5-6\% while the total importance score reduces by nearly 10\% compared to all other values for $\alpha$ considered here. We find the average width taken in the shelf of the SKUs allocated to the picking loop in this case to be smaller by about 20\%. This confirms that the importance score considers additional factors, such as the volume of the SKU (correlation coefficient of 0.52 between the width needed and the ratio between the socre and the number of picks for a SKU). In total, the analyses underline the reverse behaviour of both parts and the importance of a combination within the objective function. To conclude, $\alpha$ should be determined within the interval [0.35, 0.6]. For our ongoing analyses we fix $\alpha$ to $0.5$.

\subsubsection*{Effect of the workload balancing parameter $\delta$}
In the following, we additionally analyse the effect of the allowed deviation of picks between stations $\delta$ on the resulting objective value obtained after a runtime of one hour. Using $\alpha=0.5$, we vary the permitted relative deviation between stations in the set $\delta \in \{0.1\%, 1.0\%, 5.0\%, 10.0\%\}$ while we also analyse the results in case of ignoring balancing constraints. Table~\ref{tab:basic_delta} gives the number of SKUs included in the picking loop, the corresponding (normalised) total importance score of theses SKUs (part I of the objective function), picking efficiency (part II of the objective function), and total objective value, as well as the remaining gap after one hour runtime. In case we do not respect balancing constraints, we obtain the highest objective value with a remaining gap of 0.12\%. While this model is easy to solve, the suggested assignment is highly imbalanced with a deviation between stations of up to 16.57\%. For the settings $\delta\in\{1\%, 5\%, 10\%\}$ the objective values and gaps are very similar to each other. In contrast, the actual maximum relative deviations vary considerably and are only slightly below the corresponding permitted deviation $\delta$ in each case. At the same time, we are even able to balance the assignment on the level $\delta=0.1$. As this setting is more complex, the distance of the objective value from 1 is about twice as high as for $1\% \leq \delta \leq 10\%$ driven by a remaining gap which is also about twice as high after a runtime of one hour. From a managerial point of view, these results suggest to focus on limiting the allowed deviation to $\delta\leq 1\%$, such that the workload between stations is balanced and the risk of congestion is small, while the objective value accounting for the importance of SKUs allocated to the picking loop as well as picking efficiency reduces only slightly for equal runtimes.

\begin{table}[ht]
    \centering
    \scalebox{0.65}{
    \begin{tabular}{c|cccccc}
        $\delta$ & \# SKUs & normalised obj part I & normalised obj part II & obj value & maximum relative deviation & gap  \\
        \hline
        0.1\% & 1491 & 0.9982 & 0.9900 & 0.9949 & \,\,0.097\% & 0.63\% \\
        1.0\% & 1496 & 0.9990 & 0.9946 & 0.9973 & \,\,0.908\% & 0.35\% \\
        5.0\% & 1505 & 0.9990 & 0.9943 & 0.9971 & \,\,4.961\% & 0.37\% \\
        10.0\% & 1502 & 0.9986 & 0.9955 & 0.9974 & \,\,9.450\% & 0.33\% \\
        unrestricted & 1531 & 0.9993 & 0.9989 & 0.9991 & 16.572\% & 0.12\% \\
    \end{tabular}}
    \caption{Summary statistics on the number of SKUs included in the picking loop, the objective value, the maximum relative deviation in picks between stations and the resulting gap after a runtime of 1 hour for different values of allowed deviation $\delta$ and a fixed weighting factor $\alpha=0.5$.}
    \label{tab:basic_delta}
\end{table}

\subsubsection*{Integrated vs sequential storage assignment}
Finally, we compare our integrated model to a sequential three-stage approach, where we first solve the subproblem akin to the forward reserve allocation problem, i.e.\ the selection of SKUs to be allocated to the efficient picking loop area (part I of our objective function), with neither taking into account the picking efficiency (part II of our objective function) nor respecting balancing constraints. After a runtime of 24 seconds only the model can be solved with a gap of 0.11\%. It suggests to include 1533 SKUs into the picking loop, leading to a total (normalised) score of 0.999. Comparing these results to Table~\ref{tab:basic_delta}, we find that the score improves only slightly while we allocate 37 SKUs more to the picking loop than when also accounting for picking efficiency and limiting the deviation in picks between stations to $\delta=1\%$. At the same time, the runtime reduces comprehensively.

We then assume the set of these 1533 SKUs as given and allocate them to stations with the aim of minimising the deviation in the number of picks between stations. Within a runtime of one hour, which is sufficient to solve the integrated problem efficiently, we are not able to find a feasible solution for an assignment that satisfies a maximum deviation of 1\%. Instead, we obtain an objective value for the deviation between stations of more than 21\%. Thus, we remove those SKUs with the smallest importance score until we are able to solve the problem on the 1\% level. This holds for the 1516 SKUs with the highest importance score in the set determined before, while the corresponding total importance score decreases only slightly.

Finally, we assume the assignment of SKUs to stations as given and try to maximise the picking efficiency within the stations by determining the location in the shelves for each SKU. As each station can be optimised independently, this decomposed problem can be solved to optimality within less than 4 minutes for an individual station. We obtain a value of 0.978 for the (normalised) objective part II and, therefore, a total objective value of 0.989 when using a weighting factor $\alpha=0.5$ again. As this objective value is clearly smaller than for the integrated approach (objective value 0.997 for $\delta=1\%$), the results underline the importance of an integrated model compared to a sequential approach for the problem under consideration.

\section{Coping with short-term demand variation}
The general model developed in the previous section allows the retailer to solve the three-level storage assignment problem, i.e.\ to decide which SKUs of the assortment should be allocated to the picking loop while also determining the assignment of SKUs to stations and shelves within the warehouse. However, as illustrated by the data described in Section~\ref{sec:pikcing_data}, the demand for SKUs (and, therefore, the number of picks) is not at all constant for each day, week or even month of the year. This motivates to balance demand for each day of the week individually allowing the retailer to improve the assignment of SKUs to stations and shelves. Thus, in this section, we discuss the importance of accounting for variation in demand when assigning SKUs to shelves. We start by analysing the quality of the storage assignment determined in the previous section with respect to possible imbalance between stations on the level of days of week. Afterwards, we extend our model formulation by limiting the deviation of picks between stations on the level of days of week and compare both approaches. This allows us to derive the benefit of explicitly including variation in demand into the storage assignment decision of the retailer. As, on the other hand, the retailer needs to spend effort on data collection, data processing and computational power for the detailed analysis, this analysis forms the basis for the decision whether the benefit of the detailed solution outweighs this effort.

\subsection{Evaluating variation-agnostic storage assignments}
\label{sec:variaton_basic}
As the descriptive data analysis indicates that there exist demand patterns depending on the day of week, we now consider the degree of imbalance in the number of picks between different stations on the day-of-week level. Due to the variation in demand, two different problems may occur: (1) imbalance due to the assignment of SKUs to stations and (2) congestion driven by the assignment of SKUs to shelves within each station. 

In the basic model formulation as proposed in Section~\ref{sec:basic_model_form}, Constraints (5) and (6) limit the maximum deviation in picks between stations. In particular, the results of the basic model proposed in Section~\ref{sec:basic-model} suggest that we can efficiently limit the workload deviation at each station to up to $\delta=0.1\%$ from the average value over all stations. However, as this model only includes the average number of picks over all days of week, the deviation might be distinctly higher for single days of the week due to variation in demand. Thus, we now evaluate this allocation with the day-of-week data described in Section~\ref{sec:pikcing_data}. This allows us to analyse the quality of the basic (variation-agnostic) solution for each day of week.

\begin{table}[ht]
    \centering
    \scalebox{0.65}{
    \begin{tabular}{c|cccccc}
        day of week & Monday & Tuesday & Wednesday & Thursday & Friday & Saturday \\
        \hline        
        deviation at station 1 & \textbf{-1.25\%} & \,\,0.52\% & -0.21\% & \,\,0.09\% & -0.81\% & \textbf{\,\,1.57\%} \\ 
        deviation at station 2 & \textbf{-1.41\%} & -0.69\% & \,\,0.44\% & -0.50\% & \,\,0.36\% & \textbf{\,\,1.28\%} \\ 
        deviation at station 3 & -0.23\% & \,\,0.51\% & \textbf{-1.17\%} & -0.89\% & \,\,0.99\% & \,\,0.35\% \\ 
        deviation at station 4 & \textbf{\,\,1.01\%} & \,\,0.66\% & -0.51\% & -0.11\% & -0.20\% & -0.35\% \\ 
        deviation at station 5 & \textbf{-1.49\%} & -0.79\% & -0.14\% & \,\,0.80\% & \textbf{\,\,1.08\%} & \,\,0.60\% \\ 
        deviation at station 6 & -0.83\% & \,\,0.30\% & \,\,0.08\% & \,\,0.04\% & -0.61\% & \textbf{1.32\%} \\ 
        deviation at station 7 & \textbf{\,\,2.34\%} & -0.66\% & -0.33\% & -0.02\% & \,\,0.41\% & \textbf{-1.09\%} \\ 
        deviation at station 8 & \textbf{\,\,1.87\%} & \,\,0.15\% & \textbf{\,\,1.85\%} & \,\,0.59\% & \textbf{-1.22\%} & \textbf{-3.69\%} \\ 
        \hline
        highest negative relative deviation & \textbf{-1.49\%} & -0.79\% & \textbf{-1.17\%} & -0.89\% & \textbf{-1.22\%} & \textbf{-3.69\%} \\
        highest positive relative deviation & \textbf{\,\,2.34\%} & 0.66\% & \textbf{\,\,1.85\%} & 0.80\% & \textbf{\,\,1.08\%} & \textbf{\,\,1.57\%} \\
    \end{tabular}}
    \caption{Relative deviation in the number of picks between single stations and the average number of picks over all stations as well as corresponding minimum and maximum values. Absolute values exceeding a deviation level of 1\% are indicated in bold letters.}
    \label{tab:dev_weekday}
\end{table}

Table~\ref{tab:dev_weekday} gives an overview on the deviation in picks between each station and the average value over all stations for single days of week. Additionally, the two bottom lines indicate the maximum positive and negative deviation. Our findings show that the intended maximum deviation level of $\delta=0.1\%$ is violated at every day of week even if this constraint is satisfied when averaging over all days of week. As small deviations may only lead to minor impact on the overall efficiency, we focus on deviations exceeding 1\% indicated in bold letters. Falling below an intended level of deviation will not affect the completion time of the whole picking process within one day, but might lead to dissatisfaction of the workforce due to imbalance in the workload. In addition, there is potential to decrease the completion time compared to the solution proposed by the basic model for specific days of week. However, exceeding an intended deviation directly leads to congestion at some stations and, therefore, inefficiencies in the operational processes of the retailer, which should be avoided. As the deviation from the intended level is also high (larger than 1\%) for days with a high number of picks in total, such as Fridays (see Figure~\ref{fig:average_picks}), our results strongly advise to take into account variation in demand when determining an assignment of SKUs to stations.

\begin{figure}[ht]
    \centering
    \includegraphics[width = 0.8\textwidth]{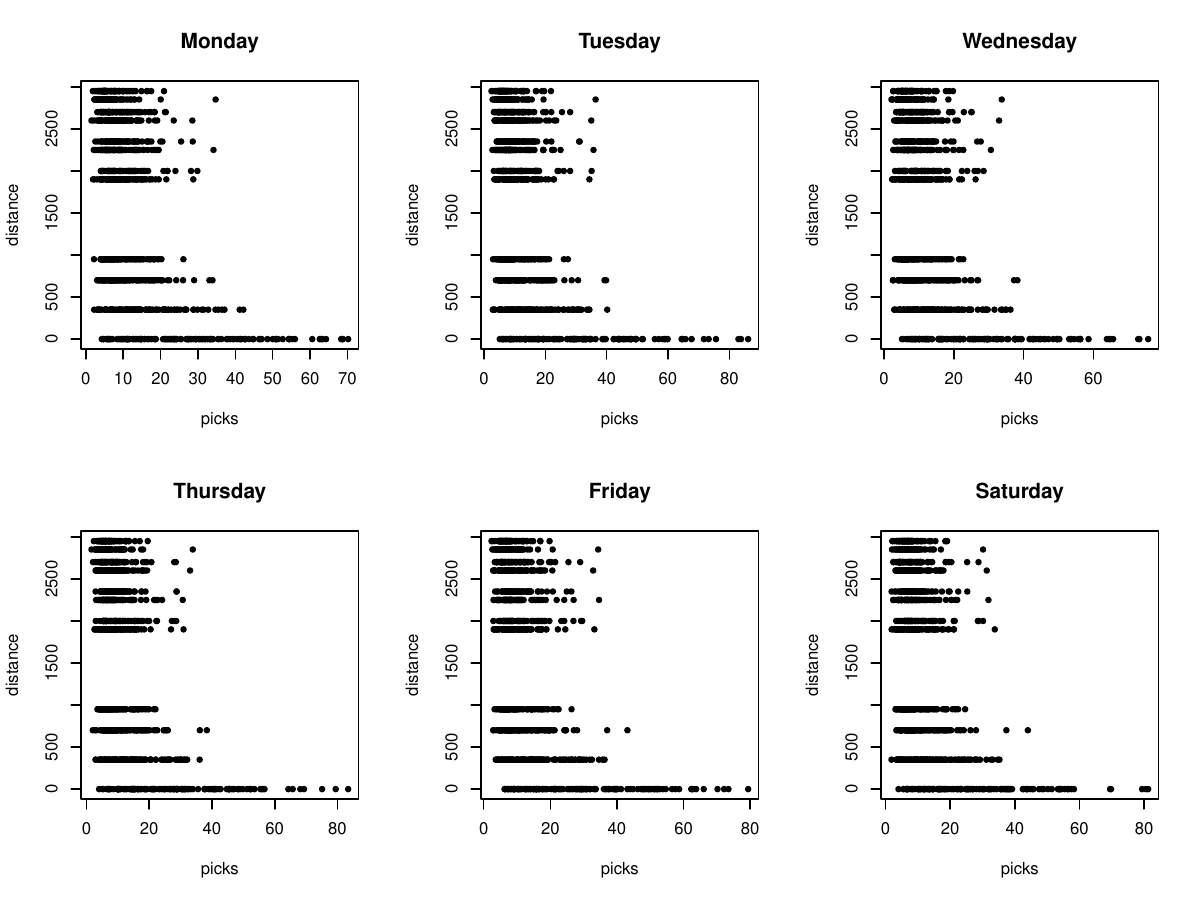}
    \caption{Allocation of SKUs and corresponding average picks per day of week to shelves with given distance to the picker for SKUs not exceeding a height of 250 mm when applying the basic model formulation with $\alpha=0.5$ and $\delta=0.1\%$.}
    \label{fig:distanz_wt_250}
\end{figure}

As the SKUs are assigned to shelves within these stations based on the average number of picks over all days of week, any variation and, therefore, a possible high relative demand at a specific day is ignored in the assignment. This may result in further congestion at this station if a SKU is assigned to an outer shelf even if there is a relatively high demand on a certain day of week. We extend the analysis illustrated in Figure~\ref{fig:distanz_vergleich} to the day-of-week level by reporting the number of picks in relation to the distance between the picker and the shelves in Figure~\ref{fig:distanz_wt_250}.\footnote{Again we exclude SKUs exceeding a height of 250mm from this figure and shown additional allocations in Figure~\ref{fig:distanz_wt_250plus} in the Appendix.} While there is variation in the figures for the different days of week, only a few outliers can be observed. As mentioned in Section~\ref{sec:basic_comp}, these outliers can be explained by their large width and also occur in the solution proposed by the basic model on the level of averages. Therefore, it can be concluded that the assignment of SKUs to shelves within stations does not lead to operational inefficiencies for specific days of week. Still, we need to address the large deviation in the number of picks between different stations on the level of days of week.

\subsection{Accounting for short-term demand variation}
Our results from the previous section indicate that it may be useful to explicitly account for day-of-week-related demand variation when determining the storage allocation for the picking loop. Indeed, we can extend the basic model from Section~\ref{sec:basic_model_form} in a way that it limits the deviation in the number of picks between different stations in the picking loop for each day of week. For this purpose, we introduce a set of days of week $t\in T$, the parameters $p_v^t$ representing the number of picks for SKU $v\in V$ and day of week $t$, and decision variables $z_k^t$ representing the workload assigned to station $k$ on day of week $t$. Analogously, we use $\delta_t$ to denote the allowed relative deviation for each day of week $t$. While in general, considering day-of-week-dependent deviations allows the retailer to consider day-of-week-specific deviation limits, in our analysis, we assume that $\delta_t$ is identical for each day of week. To enforce these maximal allowed deviations per day of week, we modify the basic model formulation from Section~\ref{sec:basic_model_form} by replacing Constraints $(4)$ -- $(6)$ with following constraints limiting the deviation on the level of days of week: \footnote{For the full model ensuring resilience against short-term demand variation as well as a table with the full notation, see Appendix \ref{sec:augmented_milp}.}

\begin{align}
     z_k^t &= \sum\limits_{v \in V}\sum\limits_{r \in R_k} p_v^t x_{v,r} \,\,\,  &&\forall k \in K  \,\,\, \forall t \in T \\
    z_k^t & \leq (1+\delta_t) \cdot \frac{1}{|K|} \sum\limits_{k \in K} z_k^t \,\,\,  &&\forall k \in K \,\,\, \forall t \in T \\
   z_k^t & \geq   (1- \delta_t) \cdot \frac{1}{|K|}  \sum\limits_{k \in K} z_k^t    \,\,\,  && \forall k \in K  \,\,\, \forall t \in T
\end{align}

\subsection{Computational experiments with the variation-aware model}

The analysis in Table~\ref{tab:dev_weekday} reveals a maximum deviation between the number of picks within one station and the average over all stations of more than 3\%. Therefore, we first test whether the extended and more complex model formulation, ensuring for resilience against short-term demand variation, allows us to limit the deviation on the day of week level to $\delta_t = 3\%$. After a runtime of less than 25 minutes the model can be solved with a (normalised) objective value of 0.9946 (compared to 0.9982 for the basic model with $\delta = 0.1\%$) and a remaining gap of 0.58\%. These results do not change over time when allowing for a runtime of up to one hour. We then limit the deviation on the day of week level to $\delta_t = 1\%$. Here, we reach a remaining gap of 0.53\% after about four hours with an objective value of 0.9951. Compared to the findings for the model with $\delta_t = 3\%$ the slightly higher objective value can be explained by the slightly smaller gap, while the runtime until reaching these results is about ten times as high. However, for $\delta_t = 0.1\%$ we do not find a feasible solution within three hours. This suggests that there are no promising solutions in terms of the objective value under such strong restrictions.

\begin{table}[ht]
    \centering
    \scalebox{0.65}{
    \begin{tabular}{r|cc|cc}
        & \multicolumn{2}{c}{highest negative relative deviation}
         & \multicolumn{2}{c}{highest positive relative deviation} \\
        day of week & Basic model & Extended model & Basic model & Extended model\\
        \hline
        Monday & -1.494\% & \textbf{-0.836\%} & 2.339\% & \textbf{0.972\%} \\
        Tuesday & -0.792\% & \textbf{-0.571\%} & 0.664\% & 0.800\% \\
        Wednesday & -1.173\% & \textbf{-0.843\%} & 1.850\% & \textbf{0.417\%} \\
        Thursday & -0.887\% & \textbf{-0.798\%} & 0.801\% & 0.879\% \\
        Friday & -1.224\% & \textbf{-0.999\%} & 1.082\% & \textbf{0.679\%} \\
        Saturday & -3.687\% & \textbf{-0.701\%} & 1.572\% & \textbf{0.955\%} \\
    \end{tabular}}
    \caption{Minimum and maximum relative deviation in the number of picks between single stations and the average number of picks over all stations for the basic model (see Table~\ref{tab:dev_weekday}) and the extended model. Absolute values which are smaller under the extended model are indicated in bold letters.}
    \label{tab:dev_wd_comp}
\end{table}

Table~\ref{tab:dev_wd_comp} compares the minimum and maximum relative deviation in the number of picks between stations for each day of week under the basic and the extended model. We are able to decrease the absolute values of the highest negative deviations for each day of week. At the same time, the highest positive deviations reduce for four out of six days while we find an increase of up to 20\% for the other two days. However, the positive deviations do not exceed 0.88\%. Compared to the highest positive value under the basic model (2.34\% on Mondays) this means a reduction of more than 60\%. 

To conclude, we are able to reduce the deviation between stations on the level of day of week from more than 3\% to less than 1\% when applying the variation-aware model with additional constraints. On the other hand, we obtain that the objective value barely reduces by 0.0036 and an increase in the runtime until reaching a comparable gap to that revealed by the basic model formulation for $\delta=0.1\%$ (see Table~\ref{tab:basic_delta}) from one to four hours. As we address a tactical problem of the retailer, which has not to be solved regularly, the benefit of reducing congestion in the picking loop outweighs the increased runtime. In most cases, retailers also benefit from the comprehensive reduction in the deviation for most days of week which again outweighs the slight reduction in the objective value.

\section{Conclusion}
In this paper, we develop an integrated approach to address a three-level storage assignment problem arising in a fulfilment centre operated by a leading European e-grocery retailer. The fulfilment centre can be characterised as a hybrid warehouse combining a highly efficient, partially automated picking loop with a less efficient picker-to-parts area. While the demand for e-groceries increased during recent years, at the same time the market became more competitive. This requires e-grocery retailers to improve their operational efficiency. A key challenge is the assignment of SKUs to shelves within a fulfilment centre. We optimise a bi-objective value function of the retailer by accounting for the importance of SKUs allocated to the high efficient picking loop, while also addressing the picking efficiency depending on the distance between a picker and the shelves. To avoid congestion within the picking loop, we additionally limit the permitted relative deviation in the number of picks between different stations in the constraints of our proposed optimisation model.

Our results show that we can efficiently solve the model with a remaining gap of less than 0.7\% within one hour in most settings. As we cover a tactical problem of the retailer, which has not to be solved regularly but only in case of major changes in the assortment of the retailer or customer preferences, such runtimes are reasonable and can be even extended. In addition, we can show that our integrated approach is superior compared to solving the allocation to the picking loop and the assignment to stations and shelves sequentially. 

An additional challenge addressed in this paper is the presence of day-of-week-dependent demand variation for certain SKUs. In our business case, the variation of demand is particularly high at the beginning of a week and right before the weekend. In a set of experiments, we show that a storage assignment that is based on day-of-week-agnostic average demand figures tends to exhibit a highly imbalanced workload on certain days of the week. In order to mitigate this problem, we extend the aforementioned storage assignment model to account for day-of-week-dependent demand variation. It turns out that the resulting model yields storage assignments that satisfy the imposed workload balance requirements on each day of week without compromising the quality of the solutions in terms of the (efficiency-oriented) objective value.

Future work may include refinements such as individual levels of permitted deviation between stations depending on the total number of picks on a certain day of week, that is, varying $\delta_t$ with respect to the day of week $t\in T$. As possible congestion is more critical at a day of week with high workload in total, this extension to the model can further reduce operational inefficiencies for e-grocery retailers. Furthermore, our approach is based on average day-of-week demands and does not consider between-week variations that may be partly attributed to randomness and partly to long-term effects such as seasonality or changing customer preferences. While random demand variation might be addressed by employing robust or stochastic optimisation approaches, the availability of data spanning multiple years would allow to detect structural long-term demand changes that justify a re-arrangement of the storage assignment. Finally, while this paper deals with an existing fulfilment centre, future research could also consider the strategic problem of designing warehouses, i.e.\ include the decision on the size of the picking loop, the number of stations and the shape of shelves.

\bibliographystyle{apalike} 
\bibliography{library}

\newpage
\appendix
\section{Appendix}
\subsection{Representation of shelves}
\begin{figure}[ht]
    \centering
    \includegraphics[width=0.4\textwidth]{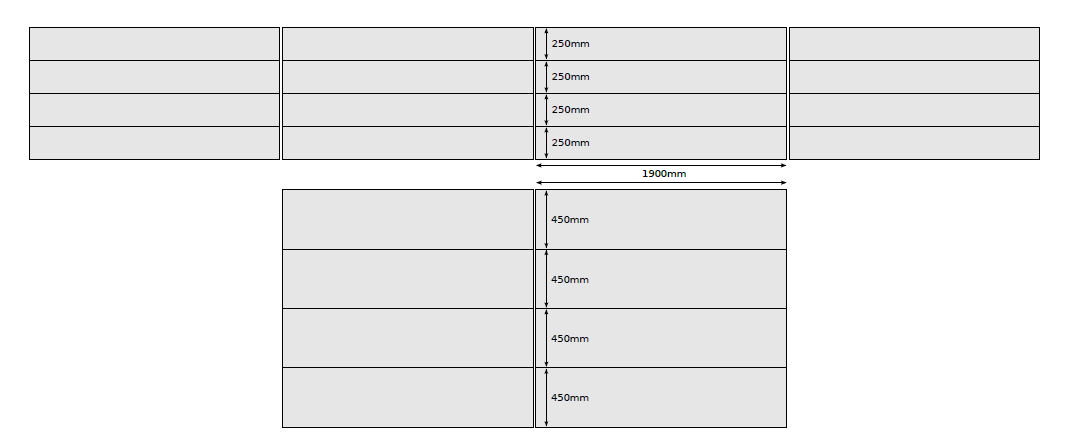}
    \caption{Representation of the structure of shelves}
    \label{fig:shelves}
\end{figure}

\subsection{Boxplot on the number of units per order line}
\begin{figure}[ht]
    \centering
    \includegraphics[width=0.4\textwidth]{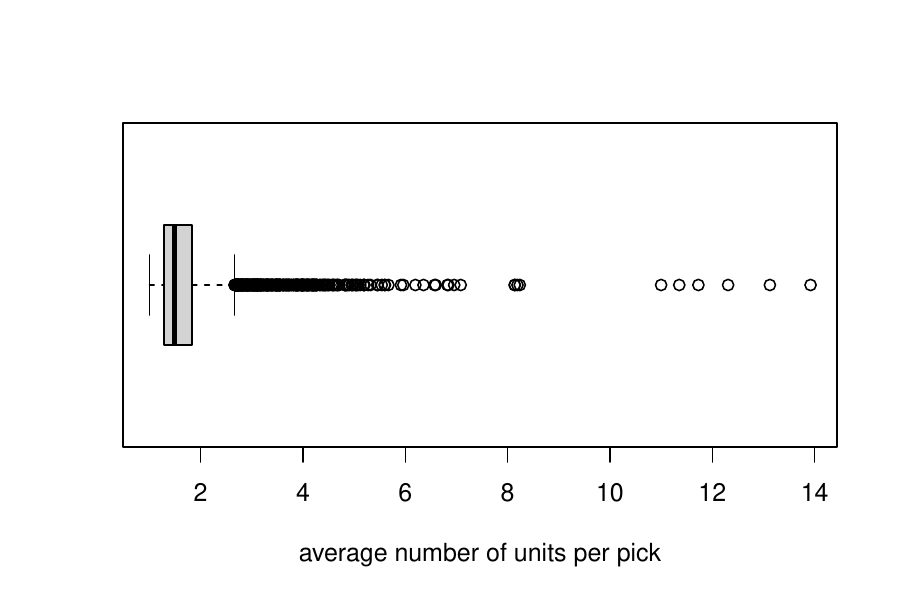}
    \caption{Boxplot of the average number of units within one order line for the SKUs in the assortment of the retailer appropriate for the picking loop.}
    \label{fig:units_pick}
\end{figure}

\newpage
\subsection{MILP Formulation for the integrated three-level storage assignment}
\begin{table}[ht]
    \centering
    \scalebox{0.85}{
    \begin{tabular}{cl}
        \hline
        \textbf{Sets} & \\
        $V$ & Set of SKUs.\\
        $K$ & Set of stations in the picking loop, ordered and indexed\\
         & by integers ($K=\{1,\ldots,|K|\}$.\\
        $R$ & Set of shelves in the picking loop with $R_k$ denoting the subset of shelves\\
         & at station $k\in K$ and $R^v$ corresponding to the subset of shelves $R$\\
         & that can fit SKU $v\in V$.\\
        $O$ & Set of precedence order ranks with $O=\{1,\ldots,\vert O\vert \}$,\\ 
         & where $o_v\leq o_{v'}$ iff $v$ has to be assigned to an earlier station than $v'$.\\
        \hline
        \textbf{Parameters} & \\
        $s_v$ & Importance score of SKU $v\in V$ \\
        $h_v$ & Height taken in the shelf SKU $v\in V$ \\
        $w_v$ & Width taken in the shelf by the target stock of SKU $v\in V$ \\
        $k_r$ & Station $k\in K$ shelf $r \in R$ is located in.\\
        $z_k$ & Workload at station $k\in K$. The average over all stations is denoted by $z$.\\
        $\delta$ & Threshold denoting the maximum permitted relative deviation of\\
         & workload $z_k$ of a station $k\in K$ from the average $z$ over all stations\\
        $h_r$ & height of shelve $r\in R$.\\
        $w_r$ & width of shelve $r\in R$.\\
        $d_r$ & Distance between the picker and shelve $r\in R$ with picking efficiency $\frac{1}{d_r}$.\\
        $g$ & Minimum distance between each two SKUs stored next to each other.\\
        \hline
        \textbf{Decision variables} & \\
        $x_{v,r}\in\{0,1\}$ & 1 iff SKU $v\in V$ is assigned to shelve $r\in R$.\\
        $y_o$ & Last station to which a SKU with order rank $o\in O$ can be assigned.\\
        \hline
    \end{tabular}}
    \caption{Table of set, parameters, and decision variables for the MILP.}
    \label{MILP_basic}
\end{table}

\vspace{-0.5cm}

\begin{equation*}
        \max \frac{\alpha}{\gamma_1} \underbrace{\sum\limits_{v \in V}\sum\limits_{r \in R^v} s_v x_{v,r}}_{\substack{I}} + \frac{1-\alpha}{\gamma_2} \underbrace{ \sum\limits_{v \in V}\sum\limits_{r \in R^v} \frac{1}{d_r} p_v x_{v,r}}_{\substack{II}}
\end{equation*}

\noindent
\begin{align*}
        \sum\limits_{r\in R^v} x_{v,r} & \leq 1 \,\,\, &&\forall \,\,\, v\in V \\
        k_r  x_{v,r} &\leq y_{o} \,\,\, &&\forall \,\,\, o \in O, v \in V^{o}, r \in R^{v} \\
        k_r  x_{v,r}  &\geq y_{o-1} \,\,\, &&\forall \,\,\, o  \in O  \setminus \{ 1 \}, v \in V^{o}, r \in R^{v} \\
        z_k &= \sum\limits_{v \in V}\sum\limits_{r \in R_k} p_v x_{v,r} \,\,\,  &&\forall \,\,\, k \in K \\
        z_k & \leq (1+\delta) \cdot \frac{1}{|K|} \sum\limits_{l \in K} z_l \,\,\,  &&\forall \,\,\, k \in K  \\
        z_k & \geq   (1- \delta) \cdot \frac{1}{|K|}  \sum\limits_{l \in K} z_l    \,\,\,  && \forall \,\,\, k \in K  \\
        w_r & \geq \sum\limits_{v \in V} (w_v + g) \cdot x_{v,r} - g \,\,\, &&\forall \,\,\, r\in R\\
        x_{v,r} & \in  \{0, 1\} \,\,\, &&\forall \,\,\, v \in V, r \in R^v  \\
        y_{o} & \in \{1,...,|K|\} \,\,\,& &\forall \,\,\, o \in O
\end{align*}

\newpage
\subsection{Augmented MILP formulation accounting for demand variation in the storage assignment.} \label{sec:augmented_milp}
\begin{table}[htp!]
    \centering
    \scalebox{0.85}{
    \begin{tabular}{cl}
        \hline
        \textbf{Sets} & \\
        $V$ & Set of SKUs.\\
        $K$ & Set of stations in the picking loop, ordered and indexed\\
         & by integers ($K=\{1,\ldots,|K|\}$.\\
        $R$ & Set of shelves in the picking loop with $R_k$ denoting the subset of shelves\\
         & at station $k\in K$ and $R^v$ corresponding to the subset of shelves $R$\\
         & that can fit SKU $v\in V$.\\
        $O$ & Set of precedence order ranks with $O=\{1,\ldots,\vert O\vert \}$,\\ 
         & where $o_v\leq o_{v'}$ iff $v$ has to be assigned to an earlier station than $v'$.\\
        $T$ & Set of days of week.\\
        \hline
        \textbf{Parameters} & \\
        $s_v$ & Importance score of SKU $v\in V$ \\
        $h_v$ & Height taken in the shelf SKU $v\in V$ \\
        $w_v$ & Width taken in the shelf by the target stock of SKU $v\in V$ \\
        $k_r$ & Station $k\in K$ shelf $r \in R$ is located in.\\
        $z_k$ & Workload at station $k\in K$. The average over all stations is denoted by $z$.\\
         & The workload at station $k\in K$ at day of week $t\in T$ is denoted by $z_k^t$.\\
        $\delta$ & Threshold denoting the maximum permitted relative deviation of\\
         & workload $z_k$ of a station $k\in K$ from the average $z$ over all stations.\\
         & The threshold at day of week $t\in T$ is denoted by $\delta_t$.\\
        $h_r$ & height of shelve $r\in R$.\\
        $w_r$ & width of shelve $r\in R$.\\
        $d_r$ & Distance between the picker and shelve $r\in R$ with picking efficiency $\frac{1}{d_r}$.\\
        $g$ & Minimum distance between each two SKUs stored next to each other.\\
        \hline
        \textbf{Decision variables} & \\
        $x_{v,r}\in\{0,1\}$ & 1 iff SKU $v\in V$ is assigned to shelve $r\in R$.\\
        $y_o$ & Last station to which a SKU with order rank $o\in O$ can be assigned.\\
        \hline
    \end{tabular}}
    \caption{Table of set, parameters, and decision variables for the MILP.}
    \label{MILP_extended}
\end{table}

\begin{equation*}
        \max \frac{\alpha}{\gamma_1} \underbrace{\sum\limits_{v \in V}\sum\limits_{r \in R^v} s_v x_{v,r}}_{\substack{I}} + \frac{1-\alpha}{\gamma_2} \underbrace{ \sum\limits_{v \in V}\sum\limits_{r \in R^v} \frac{1}{d_r} p_v x_{v,r}}_{\substack{II}}
\end{equation*}

\noindent
\begin{align*}
        \sum\limits_{r\in R^v} x_{v,r} & \leq 1 \,\,\, && \forall \,\,\, v\in V \\
        k_r  x_{v,r} &\leq y_{o} \,\,\, &&\forall \,\,\, o \in O, v \in V^{o}, r \in R^{v} \\
        k_r  x_{v,r}  &\geq y_{o-1} \,\,\, &&\forall \,\,\, o  \in O  \setminus \{ 1 \}, v \in V^{o}, r \in R^{v} \\
        z_k^t &= \sum\limits_{v \in V}\sum\limits_{r \in R_k} p_v^t x_{v,r} \,\,\,  && \forall \,\,\, k \in K , t \in T \\
        z_k^t & \leq (1+\delta_t) \cdot \frac{1}{|K|} \sum\limits_{k \in K} z_k^t \,\,\,  &&\forall \,\,\, k \in K , t \in T \\
        z_k^t & \geq   (1- \delta_t) \cdot \frac{1}{|K|}  \sum\limits_{k \in K} z_k^t \,\,\, && \forall \,\,\, k \in K , t \in T\\
        w_r & \geq \sum\limits_{v \in V} (w_v + g) \cdot x_{v,r} - g \,\,\, &&\forall \,\,\, r\in R\\
        x_{v,r} & \in  \{0, 1\} \,\,\, &&\forall \,\,\, v \in V, r \in R^v  \\
        y_{o} & \in \{1,...,|K|\} \,\,\,& &\forall \,\,\, o \in O
\end{align*}\\

\newpage

\subsection{Shelf distance of SKUs with height exceeding 250 mm}
\begin{figure}[!htb]
    \centering
    \includegraphics[width = 0.5\textwidth]{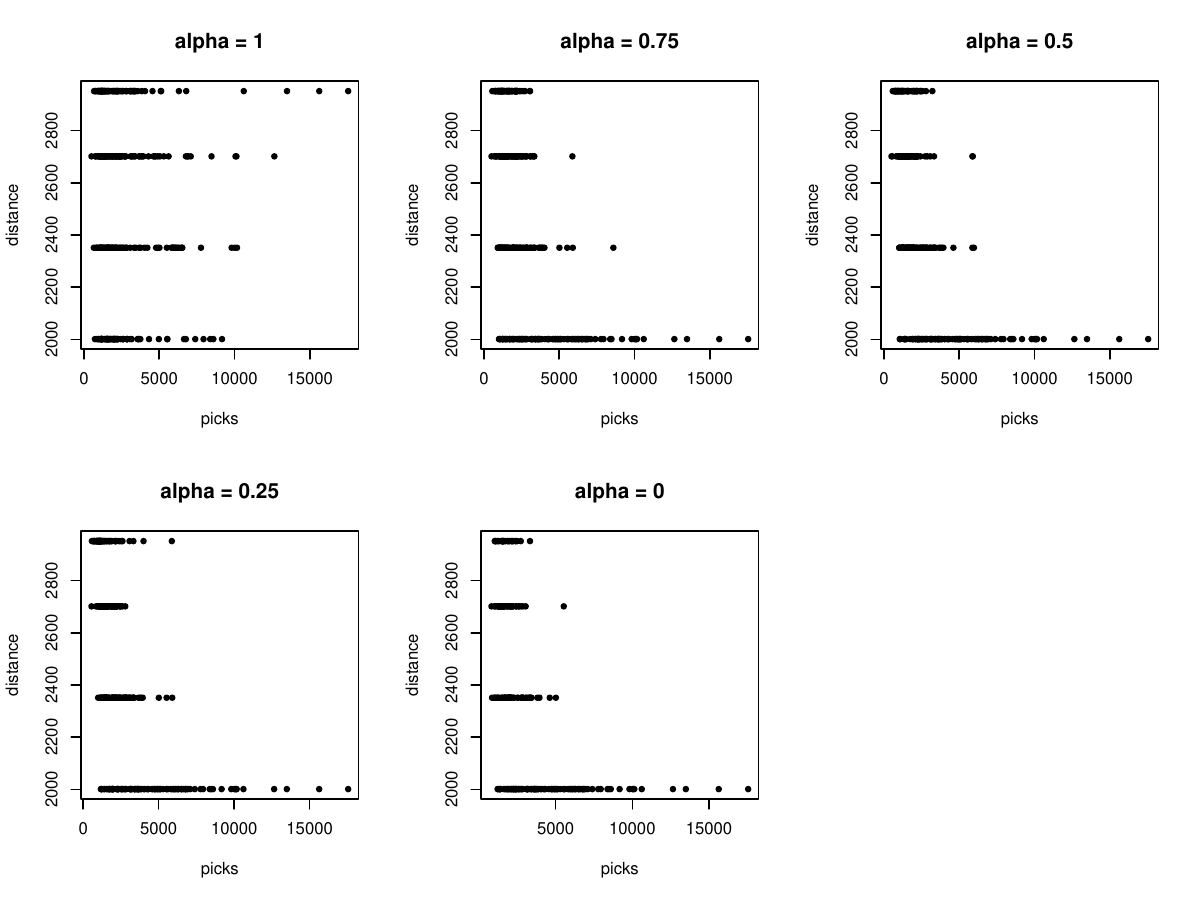}
    \caption{Allocation of SKUs and corresponding picks to shelves with given distance to the picker for different values of $\alpha$ and $\delta=1\%$. Note that the figure is limited to SKUs with a height exceeding 250 mm.}
    \label{fig:distanz_vergleich_250plus}
\end{figure}

\begin{figure}[ht]
    \centering
    \includegraphics[width = 0.5\textwidth]{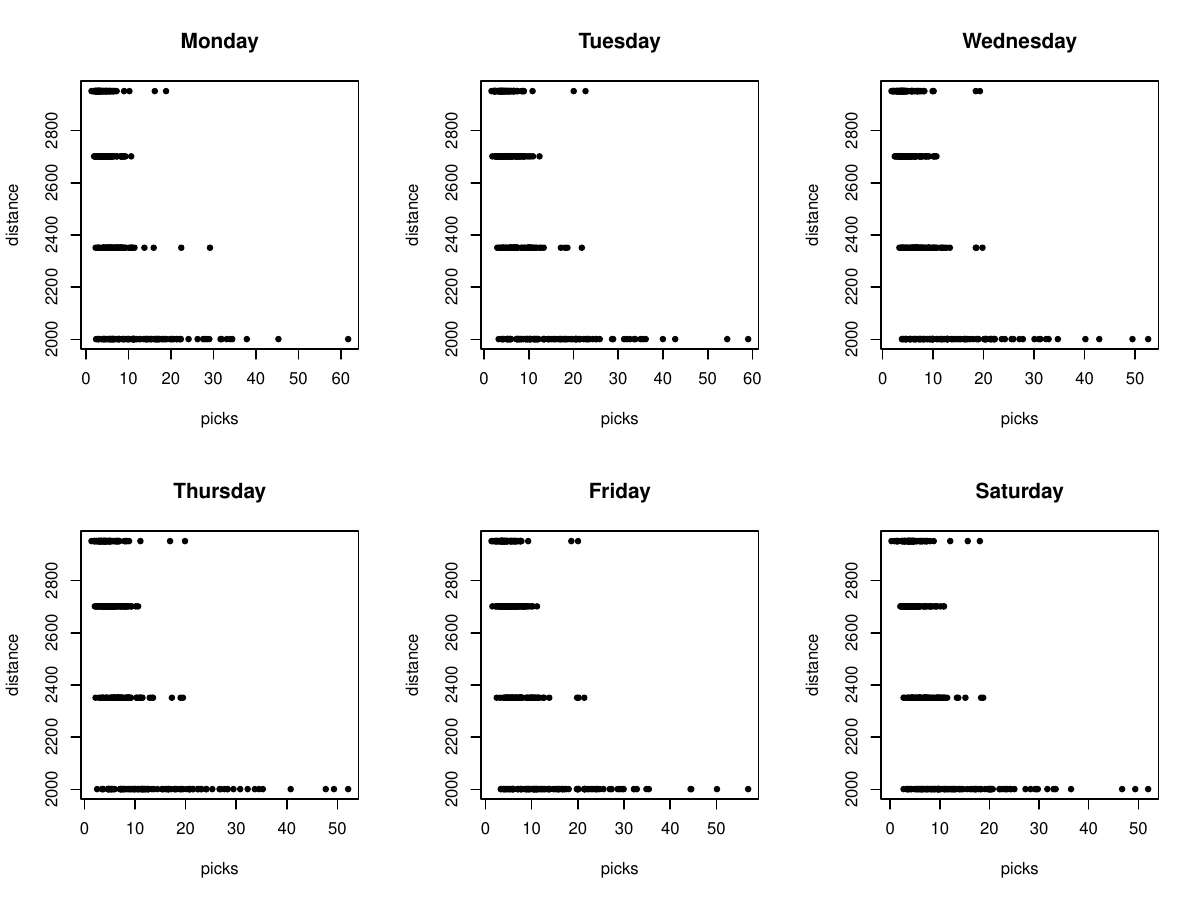}
    \caption{Allocation of SKUs and corresponding average picks per day of week to shelves with given distance to the picker for SKUs exceeding a height of 250 mm when applying the basic model formulation with $\alpha=0.5$ and $\delta=0.1\%$.}
    \label{fig:distanz_wt_250plus}
\end{figure}

\end{spacing}
\end{document}